\documentclass[a4paper,english]{elsarticle}
\usepackage[T1]{fontenc}
\usepackage[latin9]{inputenc}
\usepackage{varioref}
\usepackage{units}
\usepackage{amsmath}
\usepackage{amssymb}
\usepackage{graphicx}
\usepackage{esint}
\usepackage{caption}
\usepackage{subcaption}

\makeatletter

\providecommand{\tabularnewline}{\\}

\journal{Elsevier}
\usepackage{upgreek}
\usepackage[bookmarks,bookmarksopen,bookmarksdepth=6]{hyperref}
\usepackage{cases}
\usepackage{url}
\usepackage{lineno}

\usepackage{a4wide}

\makeatother

\usepackage{babel}
\begin{document}

\title{Measurement of the drift velocities of electrons and holes in high-ohmic
<100> silicon}

\author[]{C.~Scharf\corref{cor1}}

\author[]{R.~Klanner}

\cortext[cor1]{Corresponding author. Email address: Christian.Scharf@desy.de, Telephone:
+49 40 8998 4725}
\address{Institute for Experimental Physics, University of Hamburg, Hamburg,
Germany

\pdfbookmark[5]{Abstract}{abstract}}

\begin{abstract}
 Measurements of the drift velocities of electrons and holes as functions of electric field and temperature in high-purity n- and p-type silicon with <100> lattice orientation are presented.
 The measurements cover electric field values between 2.5 and $\unit[50]{kV/cm}$ and temperatures between 233 and $\unit[333]{K}$.
 For both electrons and holes differences of more than $\unit[15]{\%}$ are found between our <100> results and the <111> drift velocities from literature, which are frequently also used for simulating <100> sensors.
 For electrons, the <100> results agree with previous <100> measurements; however, for holes differences between 5 and $\unit[15]{\%}$ are observed for fields above $\unit[10]{kV/cm}$. Combining our results with published data of low-field mobilities, we derive parametrizations of the drift velocities in high-ohmic <100> silicon for electrons and holes for fields up to $\unit[50]{kV/cm}$, and temperatures between 233 and $\unit[333]{K}$.
 In addition, new parametrizations for the drift velocities of electrons and holes are introduced, which provide somewhat better descriptions of existing data for <111> silicon than the standard parametrization.
\end{abstract}


\maketitle
 \tableofcontents
 \newpage
 \pagenumbering{arabic}

\section{Introduction}
 \label{sect:Introduction}

 Simulations of silicon sensors  and the extraction of parameters from measurements, rely on the knowledge of a number of material parameters, for example the drift velocities of electrons and holes as functions of electric field and temperature in high-resistivity silicon.
 Presently, the information on drift velocities for silicon with <100> lattice orientation is quite limited. However, <100> silicon is widely used in high energy physics detectors as the defect concentration at the surface is small compared to <111> silicon. Therefore, results from <111> silicon are typically also used for the simulation and analysis of <100> data.
 We have investigated how well we can simulate the measured current transients in non-irradiated pad sensors built on <100> silicon and found that by using the data of Refs.~\citep{jacoboni1977review,becker2011measurements} we were unable to describe the measurements. Therefore, we determined the drift velocities of electrons and holes in silicon of <100> lattice orientation as functions of the electric field and the temperature using two different methods.

 \section{Sensors investigated and experimental setup}
  \label{sect:Experiment}

 \subsection{Sensors investigated}
  \label{subsect:Sensors}

 Three different sensors were investigated, $\text{p}^{\text{+}}$n$\text{n}^{\text{+}}$ and $\text{n}^{\text{+}}$p$\text{p}^{\text{+}}$ pad diodes, two manufactured by Hamamatsu (HPK)~\citep{HPKweb} and one by CiS~\citep{CiSweb}, on <100> float-zone silicon.
Physical parameters of the diodes are given in Table~\ref{tab:Properties-samples}.
 The doping concentration was obtained from capacitance-voltage measurements.
 The mean oxygen concentration of the diodes was $\unit[5-10\cdot10^{16}]{cm^{-3}}$.
 The active thicknesses of the diodes were obtained from the mechanical thickness, which was measured using a caliper and dielectric measurements, minus the depths of the implants.
 For the HPK diodes the depths of the $\text{n}^{\text{+}}$ and $\text{p}^{\text{+}}$ implants, determined using spreading resistance measurements by SGS Institut Fresenius GmbH~\citep{Fresenius}, were used.
 As we have no measurements of the implant depths for the CiS diode, the values of the HPK diodes were used.
 The values determined from the capacitance above full depletion were used as a cross check.
 The error of the active thickness is estimated to be $\unit[1]{\%}$.
 To allow injection of light from both sides of the pad diodes, the junction (front) side contact is a square aluminum ring, which leaves free the center of the diode, and the opposite (rear) side contact is an aluminum grid.

\begin{table}[!ht]
\begin{centering}
\begin{tabular}{|c||c|c|c|c|c|c|c|c|}
\hline
Vendor & Type & $w$ & $U_{dep}$ & $N_{eff}$ & $A$ & $C_{fd}$ & $\langle E_{min}\rangle$ & $\langle E_{max}\rangle$\tabularnewline
& & $\unit[]{[\upmu m]}$ & $\unit[]{[V]}$ & $\unit[]{[cm^{-3}]}$ & $\unit[]{[mm^{2}]}$ & $\unit[]{[pF]}$ & $\unit[]{[kV/cm]}$ & $\unit[]{[kV/cm]}$ \tabularnewline
\hline
\hline
HPK & n & $200\pm2$ & $87.5\pm2.6$ & $2.9\cdot10^{12}$ & 4.4 & 2.7 & 5.0 & 50.0 \tabularnewline
\hline
HPK & p & $200\pm2$ & $115.0\pm3.5$ & $3.8\cdot10^{12}$ & 4.4 & 2.7 & 6.0 & 50.0 \tabularnewline
\hline
CiS & n & $287\pm3$ & $48.2\pm1.4$ & $0.8\cdot10^{12}$ & 24.5 & 9.2 & 2.4 & 34.8\tabularnewline
\hline
\end{tabular}
\par\end{centering}
 \caption[Physical properties of the test structures we investigated.]{Physical properties of the investigated pad diodes; $w$ denotes the active thickness, $U_{dep}$ the depletion voltage, $N_{eff}$ the effective bulk doping, $A$ the pad area, $C_{fd}$ the capacitance above full depletion, and $\langle E_{min}\rangle$ and $\langle E_{max}\rangle$ are the minimum and maximum mean electric fields in the diodes where the drift velocity was measured.
\label{tab:Properties-samples}}
\end{table}

 \subsection{Experimental setup}
  \label{subsect:Setup}

 In order to determine the drift velocities of electrons and holes separately the transient current technique (TCT) was employed, with charge carriers produced by pico-second laser light with different wavelengths.
 Most of the measurements were performed at the Hamburg setup, which is described in more detail in Refs.~\citep{becker2010thesis,scharf2014thesis}. Additional control measurements were performed at the TCT setup at the CERN-SSD Lab~\citep{CERNSSDLAB}.
 Current transients generated by laser pulses with a $\unit[FWHM=50]{ps}$ and a pulse frequency of $\unit[200]{Hz}$ were recorded via $\unit[3]{m}$ RG58 cable, a bias-T, and an amplifier~\citep{Ampli} using a Tektronix DPO 4104 oscilloscope with a bandwidth of $\unit[1]{GHz}$ and a sampling rate of $\unit[5]{GS/s}$.
 To reduce the effects of noise 512 transients were averaged at each measurement step.

 Two lasers~\citep{Laser} with different wavelengths were used:
 a red laser with light of $\unit[\lambda=675]{nm}$ wavelength for front- and rear-side illumination, and an infrared laser with $\unit[\lambda=1063]{nm}$ for front-side illumination.
 The absorption length of $\unit[675]{nm}$ light in silicon is $\unit[\Lambda\approx3.3]{\upmu m}$ at room temperature (RT), and charge carriers are created only close to surface of the diode.
 Thus only one carrier type drifts through the whole diode and generates most of the transient current.
 For illumination of the $\text{p}^{\text{+}}\text{n}$~junction of a $\text{p}^{\text{+}}$n$\text{n}^{\text{+}}$ diode, the generated holes are collected immediately at the close-by $\text{p}^{\text{+}}$~contact, while the  electrons drift through the n bulk to the $\text{n}^{\text{+}}$~contact.
 Likewise, for the illumination of the $\text{n}^{\text{+}}$~contact the electrons are collected immediately, while the holes drift through the bulk to the $\text{p}^{\text{+}}$~contact.
 The absorption length of the infrared laser light of $\unit[\Lambda\approx1]{mm}$ at RT is much larger than the active thickness of the investigated diodes.
 Therefore, the electron-hole pairs are generated approximately uniformly along the path of the light and both electron and holes contribute equally to the current transient.

 The measurements were performed at different temperatures for bias voltages of  $\unit[U_{dep}<U_{bias}\leq 1000]{V}$ in steps of $\unit[10]{V}$.
 For the HPK n-type diode, the measurements were performed at seven temperatures: $\unit[T=(233,\,253,\,263,\,273,\,293,\,313,\,333)]{K}$\label{Temperatures}.
 For every value of $U_{bias}$ and $T$, the measurements were performed for all three laser configurations:
 front illumination with the red laser,
 rear illumination with the red laser, and
 front illumination with the infrared laser.

\section{Analysis methods}
 \label{sect:Analysis}
 \subsection{Transient current simulation}
  \label{subsect:Simulation}

 A simulation of charge transport including the diffusion of charge carriers was used to describe the TCT measurements.
 A more detailed description can be found in Refs.~\citep{becker2011measurements,becker2010thesis,scharf2014thesis}.
 The following description of the simulation is partly taken from Ref.~\citep{transferf}.
 Electron-hole pairs are generated on a grid with $\Delta x = \unit[50]{nm}$ spacing according to an exponential function with the light-attenuation length $\Lambda (T)$.
 The temperature dependence of $\Lambda (T)$ is modeled according to Ref.~\citep{kramberger2001thesis}.
 The charge carriers are drifted in time steps of $\unit[\Delta t=10]{ps}$, taking into account diffusion by a Gaussian with variance $\sigma_{e}=\sqrt{(2\cdot\mu_{e}\cdot k_{B}\cdot T/q_{0})\cdot \Delta t}$ for electrons, and similarly for holes.
 The Boltzmann constant is $k_{B}$, the absolute temperature $T$, the elementary charge $q_{0}$, and the field-dependent electron and hole mobilities $\mu_{e}$ and $\mu_{h}$.
 The current induced in the electrodes in the time interval between $i\cdot\Delta t$ and $(i+1)\cdot\Delta t$ is calculated according to $I_{i}^{sim}=q_{0}/\Delta t\cdot\sum_{j}\frac{j\cdot\Delta x}{w}\big[(N_{i+1,j}^{e}-N_{i,j}^{e})-\big(N_{i+1,j}^{h}-N_{i,j}^{h}\big)\big]$, where $N_{i,j}^{e}$ is the number of electrons and $N_{i,j}^{h}$ the number of holes at the grid point $j$ at time $i\cdot\Delta t$.
 Effects like charge trapping or charge multiplication are not considered.
 A constant space charge density $q_0 \cdot N_{eff}$ is assumed, resulting in a linear electric field.
 It has been verified, that the current transients are hardly affected if the electric field
 simulated with SYNOPSYS-TCAD~\citep{TCAD} for realistic doping distributions at the implants is used instead.

 The transfer function $R$ of the electronics circuit is determined using the convolution theorem $\mathcal{F}\{f\otimes g\}=\mathcal{F}\{f\}\cdot\mathcal{F}\{g\}$ of Fourier transforms.
 The measured TCT pulse has a sampling interval of $\unit[200]{ps}$, and a spline interpolation of the measurement $I^{int}$ is used to obtain values for the $\unit[10]{ps}$ time steps of the simulation.
 The transfer function is obtained by
$R=\mathcal{F}^{-1}\Big(\frac{\mathcal{F}\{I^{int}\}}{\mathcal{F}\{I^{sim}\}}\Big)$.
  The convolved simulated signal is then obtained by $S_{k}^{sim} = \sum_l I_{k-l}^{sim} \cdot R_l$ and compared to the measurements.
 The method is described in more detail in Ref.~\citep{transferf}.

 \subsection{Mobility parametrizations}
  \label{subsect:Mobility}

 The Caughey-Thomas (CT) parametrization~\citep{caughey1967carrier,jacoboni1977review} is generally used to describe the field dependent mobility in high-ohmic silicon
 \begin{equation}
  \mu^{CT}(E)=\frac{\mu_{0}^{CT}} {\Big(1+\big(\frac{\mu_{0}^{CT}\cdot E}{v_{sat}^{CT}}\big)^{\beta}\Big)^{1/\beta}}.
  \label{CT}
 \end{equation}
 with the low field mobility $\mu_0$, the saturation velocity $v_{sat}$, and a phenomenological parameter $\beta$.

\begin{figure}[!ht]
  \centering{}\includegraphics[width=10cm]{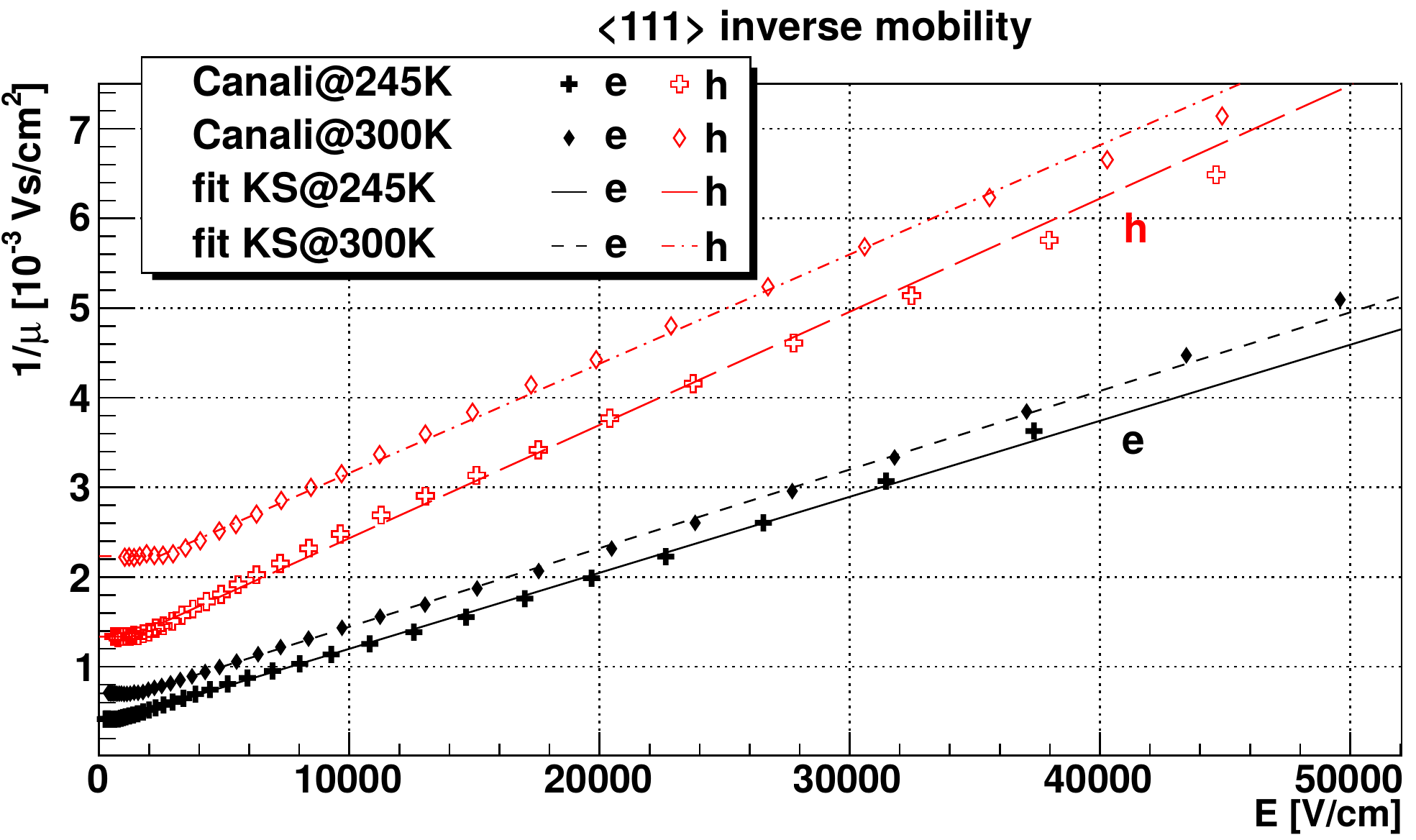}
   \caption{Inverse of the mobilities for electrons and holes as functions of the electric field for <111> silicon at $T = 245$ and 300\,K. The data have been obtained by digitizing the points of Figs.\,5 and 7 of Ref.\,\cite{canali1971drift}. The curves are fits to the data using Eq.~\ref{eq:newparlin}.
  \label{fig:canali_TOF_with_fit}}
 \end{figure}

 When comparing different mobility measurements reported in the literature, we noticed that for high electric fields $1/\mu (E)$ is to a good approximation a linear function of electric field $E$, with a very sudden transition from the constant low-field mobility $\mu _0$ to the linear $E$ dependence.
 For illustration Fig.\,\ref{fig:canali_TOF_with_fit} shows $1/\mu (E)$ for electrons and holes for <111> silicon at $T = 245$ and 300\,K obtained by digitizing figures of Ref.\,\cite{canali1971drift}, and a fit by the new parametrization (KS) presented in this paper
 \begin{eqnarray}
   1/\mu^{KS}(E) & =
   \begin{cases}
   \nicefrac{1}{\mu_{0}^{KS}} & E<E_{0}\\
   \nicefrac{1}{\mu_{0}^{KS}}+\nicefrac{1}{v_{sat}^{KS}}\cdot(E-E_{0}) & E\geq E_{0}\,.
   \end{cases}
  \label{eq:newparlin}
 \end{eqnarray}
 At $E = E_0$ the inverse mobility changes from a constant value to a linear increase.
 In order to have a function with a continuous derivative, one may use the function
 $1/\mu(E) = 1/\mu_{0}+(E - E_0) \cdot (1 + \tanh(E - E_0))/(2\,v_{sat})$,
 with the units of V/m for $E$.
 The maximum deviation of this equation from Eq.~\ref{eq:newparlin} is $\unit[10^{-3}]{\%}$.
 We note that for $E_0 = 0$ the new parametrization has the same field dependence as the Trofimenkoff (Tr) parametrization~\citep{trofimenkoff1965field}:
 \begin{eqnarray}
   \mu^{Tr}(E) & = & \frac{\mu_{0}^{Tr}}{1+\frac{\mu_{0}^{Tr}}{v_{sat}^{Tr}}\cdot E}.
  \label{eq:munewe}
\end{eqnarray}

 The physics motivation for the parametrizations is:
 at low electric fields the drift velocity $v_d$ is small compared to the thermal velocity $v_{th}$.
 Assuming an effective scattering rate $1/\tau _{\,lattice}$, the drift velocity
 $v_d(E) = \int _0 ^\infty \big((q_0 \cdot E/m^*) \cdot e^{-t/\tau _{lattice}} \big)\,\mathrm{d}t = \big(q_0 \cdot \tau_{lattice}/m^*\big)\cdot E$,
 and the low-field mobility is constant: $\mu_0 = q_0 \cdot \tau_{lattice}/m^*$, with
 the effective mass of the charge carriers $m^*$.
 At high electric fields the charge carriers can acquire an energy $\epsilon_{emission}$, which is sufficient for phonon emission.
 The phonon emission rate $1/\tau_{emission}(E)$, which increases with increasing electric field, is responsible for the additional term $\propto E$ in Eq.~\ref{eq:newparlin} leading to saturation of the drift velocity. We introduce a threshold field $E_0$ for the phonon emission.
 
 We refer to the Appendix\,1 for a comparison of the fit results for the two mobility parametrizations to the electron and hole data for <111> silicon of  Ref.~\citep{canali1971drift}.
 Here we only mention that at temperatures above $\unit[245]{K}$ the new parametrization provides a somewhat better description of the data.
 At lower temperatures none of the parametrizations provides a good description.
 For the quantitative comparison we use the root-mean-squared relative deviation
  \begin{eqnarray}
  rdev & = & \frac{1}{n}\cdot\sqrt{\sum_{i=1}^{n}\Bigg(\frac{\mu^{model} (E_{i}) - \mu^{data}(E_{i})}{\mu^{data} (E_{i})}\Bigg)^{2}}.
  \label{eq:rdev}
 \end{eqnarray}
 As an example, for the data at $T = 300$\,K the values of $rdev$ using the CT parametrization are 0.41\,\% for electrons and 0.42\,\% for holes, whereas they are 0.17\,\% and 0.28\,\% for the KS parametrization.

 For the hole mobility the quality of the description of our data could be improved significantly by adding a quadratic term to the parametrization of the inverse mobility of Eq.~\ref{eq:newparlin}:
 \begin{eqnarray}
  1/\mu_{h}(E) & = \begin{cases}
  \nicefrac{1}{\mu_{0}^{h}} & E<E_{0}\\
  \nicefrac{1}{\mu_{0}^{h}}+b\cdot(E-E_{0})+c\cdot(E-E_{0})^{2} & E\geq E_{0}.
  \end{cases}\label{eq:munewh}
 \end{eqnarray}

 The parameters of the $KS$ mobility model were determined for the temperatures at which the data were taken, as well as for the entire temperature range. For the parameter $c$ no temperature dependence has been assumed.
 The temperature dependencies of the parameters was modeled similar to Refs.~\citep{jacoboni1977review,becker2011measurements,selberherr1990evolution} as a power law
 \begin{eqnarray}
   par_{i}(T) & = & par_{i}(\unit[T=300]{K})\cdot \Big(\frac{\unit[T]{[K]}} {\unit[300]{K}}\Big)^{\alpha_{i}},
  \label{eq:parT}
 \end{eqnarray}
 with $par_{i}(\unit[T=300]{K})$ denoting the parameter values at 300\,K and the power $\alpha_{i}$.

 \subsection{Fit method}
  \label{subsect:Fit}

 For determining the drift velocities of electrons and holes the quantity
 \begin{eqnarray}
  \chi ^2 & = &const \cdot \sum_{i=1}^{n}\big(S_i^{sim} - I_i^{meas}\big)^{2},
  \label{eq:Chifit}
 \end{eqnarray}
 was minimized, where $I_i^{meas}$ are the measured current values, and $S_i^{sim}$ the simulated ones after convolution with the transfer function $R$.
 As discussed in Section\,\ref{sect:Results}, fits have been made for different data sets.
 For each sensor the transfer function $R$, determined from the measurements taken with the infrared laser at $\unit[T=313]{K}$ and $\unit[U_{bias}=1000]{V}$, was used.
 For determining $R$ the drift velocities for electrons and holes have to be known. Therefore, $R$ had to be determined in an iterative way: for a given set of measurements $R$ has been determined using the Fourier transform method described in Ref.~\citep{transferf}. Using this transfer function a set of mobility parameters has been obtained by a fit of the entire data set for a given sensor, meaning all transients at bias voltages above depletion at all temperatures for all three laser configurations. This procedure was repeated until stable results have been reached after about 3 iterations.
 It has been checked that the final results do not depend on the starting values of the mobility parameters.

\begin{figure}[!ht]
  \centering
   \begin{subfigure}[a]{7.5cm}
    \includegraphics[width=7.5cm]{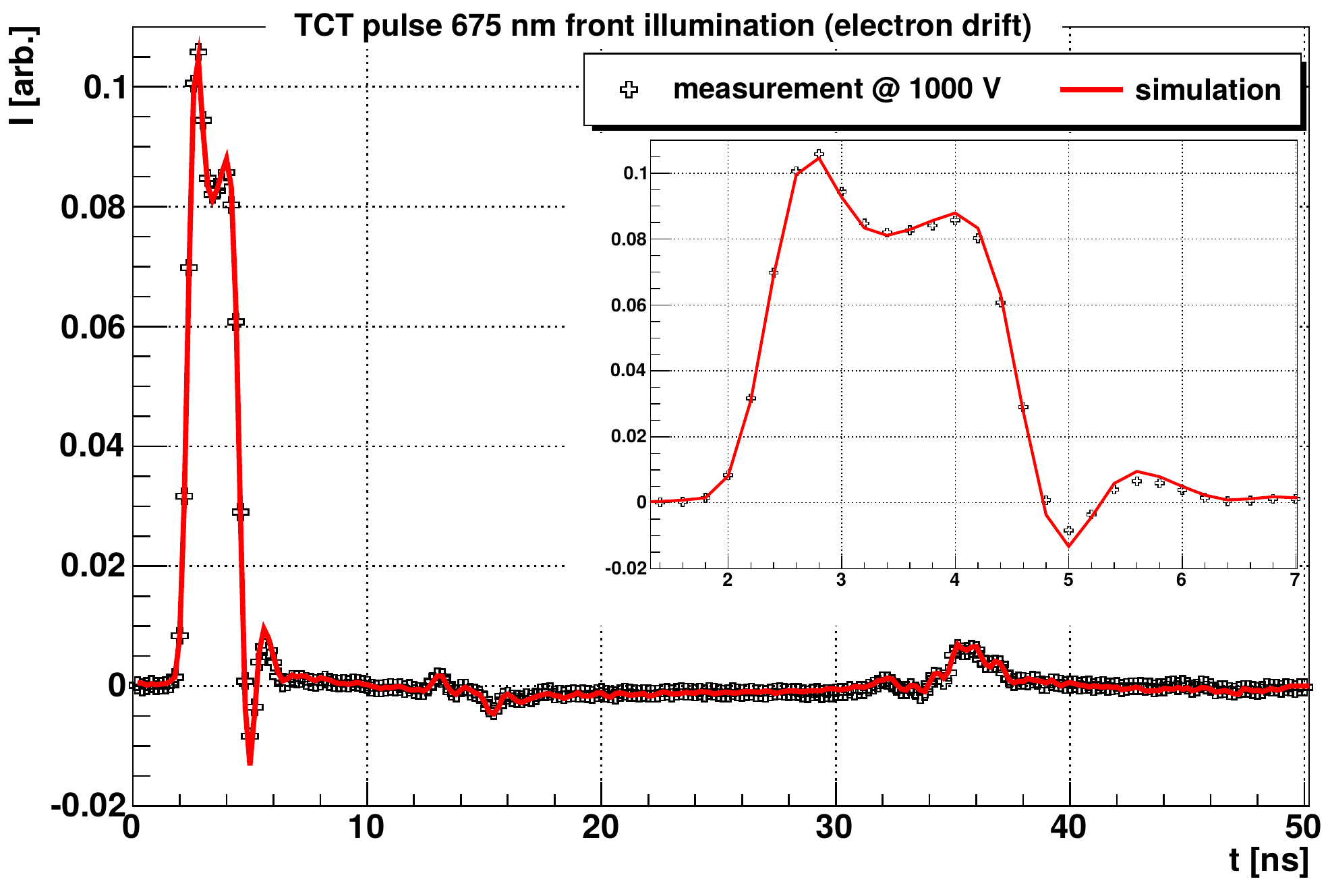}
    \caption{ }
     \label{fig:E1000}
   \end{subfigure}%
    ~
   \begin{subfigure}[a]{7.5cm}
    \includegraphics[width=7.5cm]{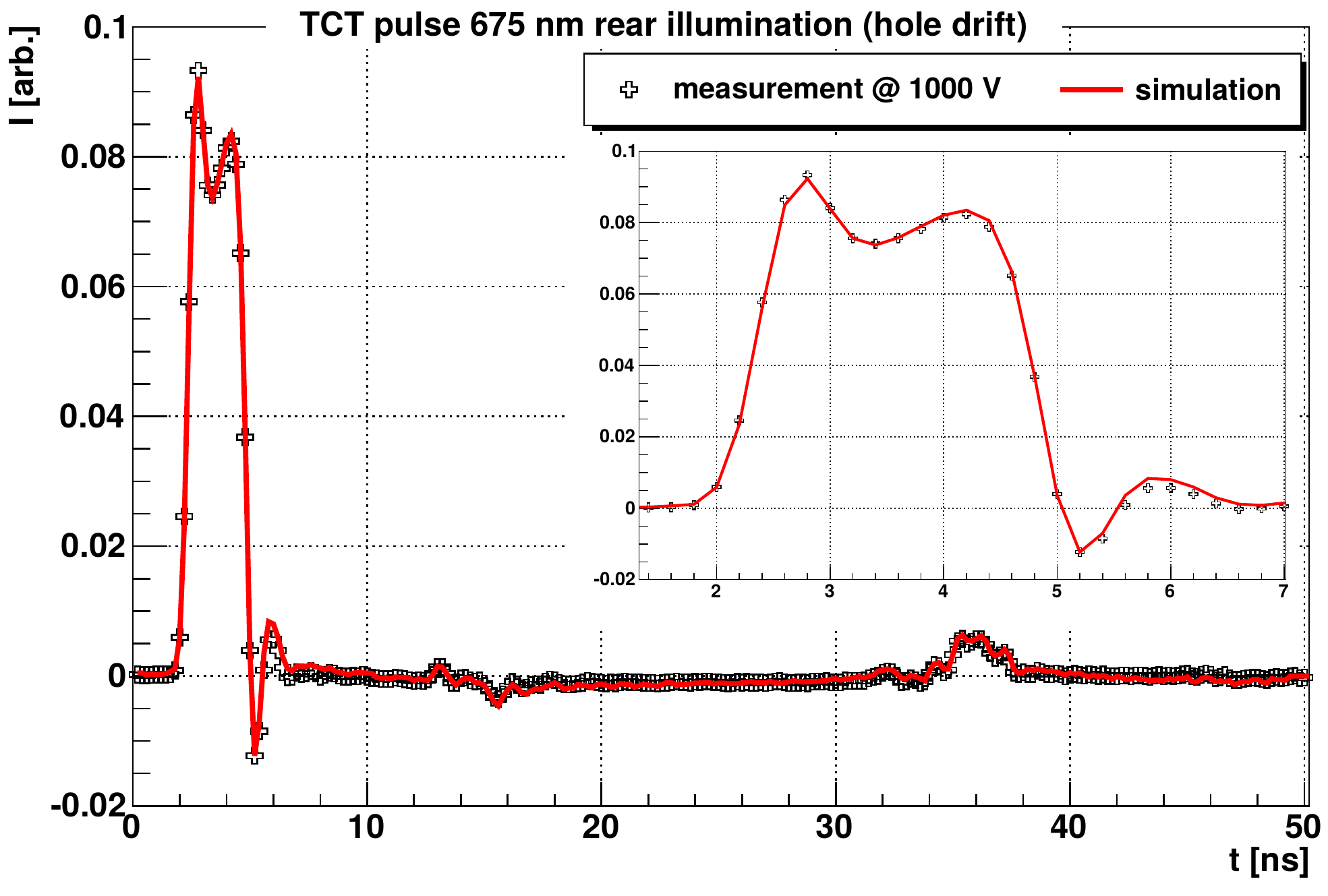}
    \caption{ }
     \label{fig:H1000}
   \end{subfigure}
    ~
   \begin{subfigure}[a]{7.5cm}
    \includegraphics[width=7.5cm]{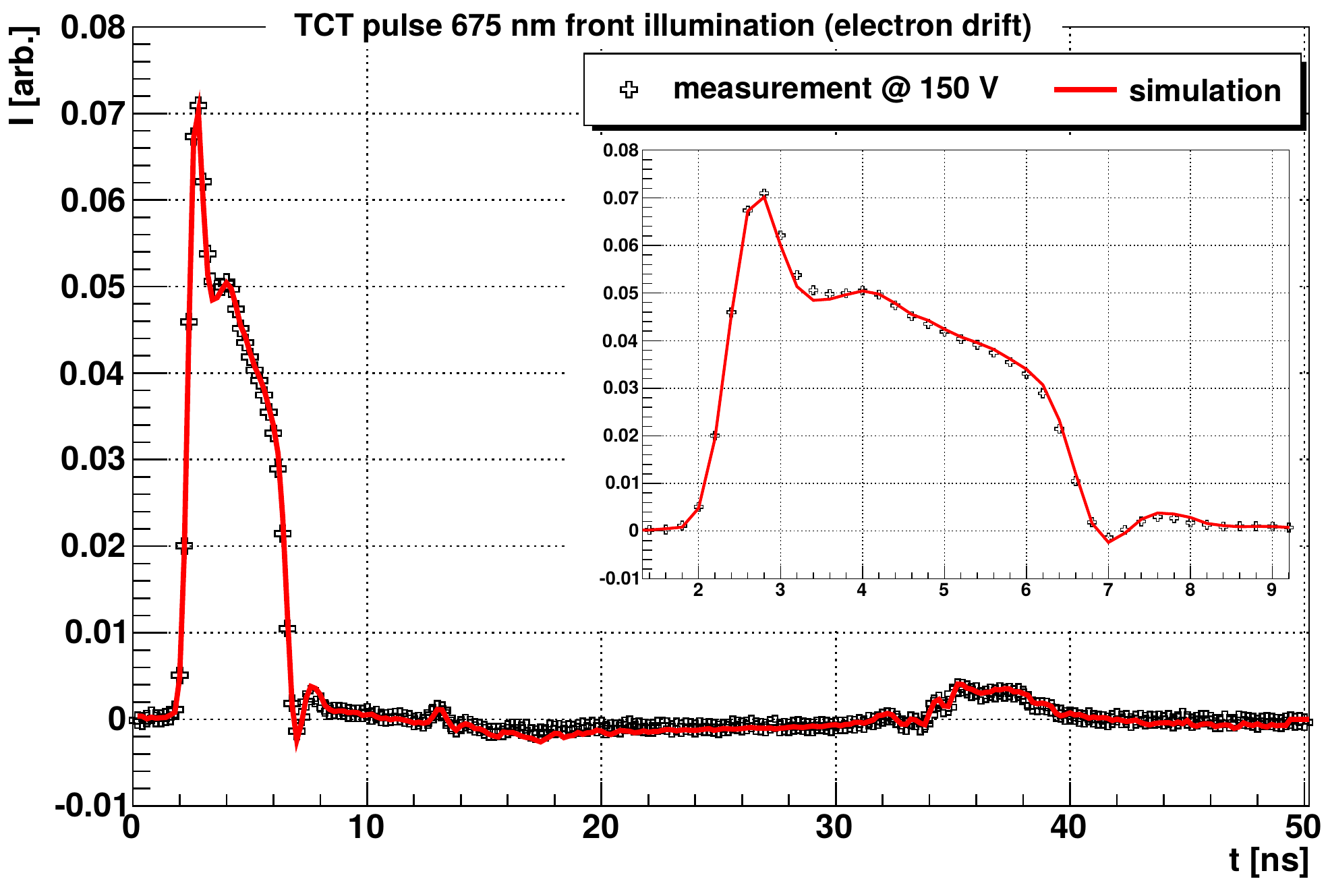}
    \caption{ }
     \label{fig:E150}
   \end{subfigure}%
    ~
   \begin{subfigure}[a]{7.5cm}
    \includegraphics[width=7.5cm]{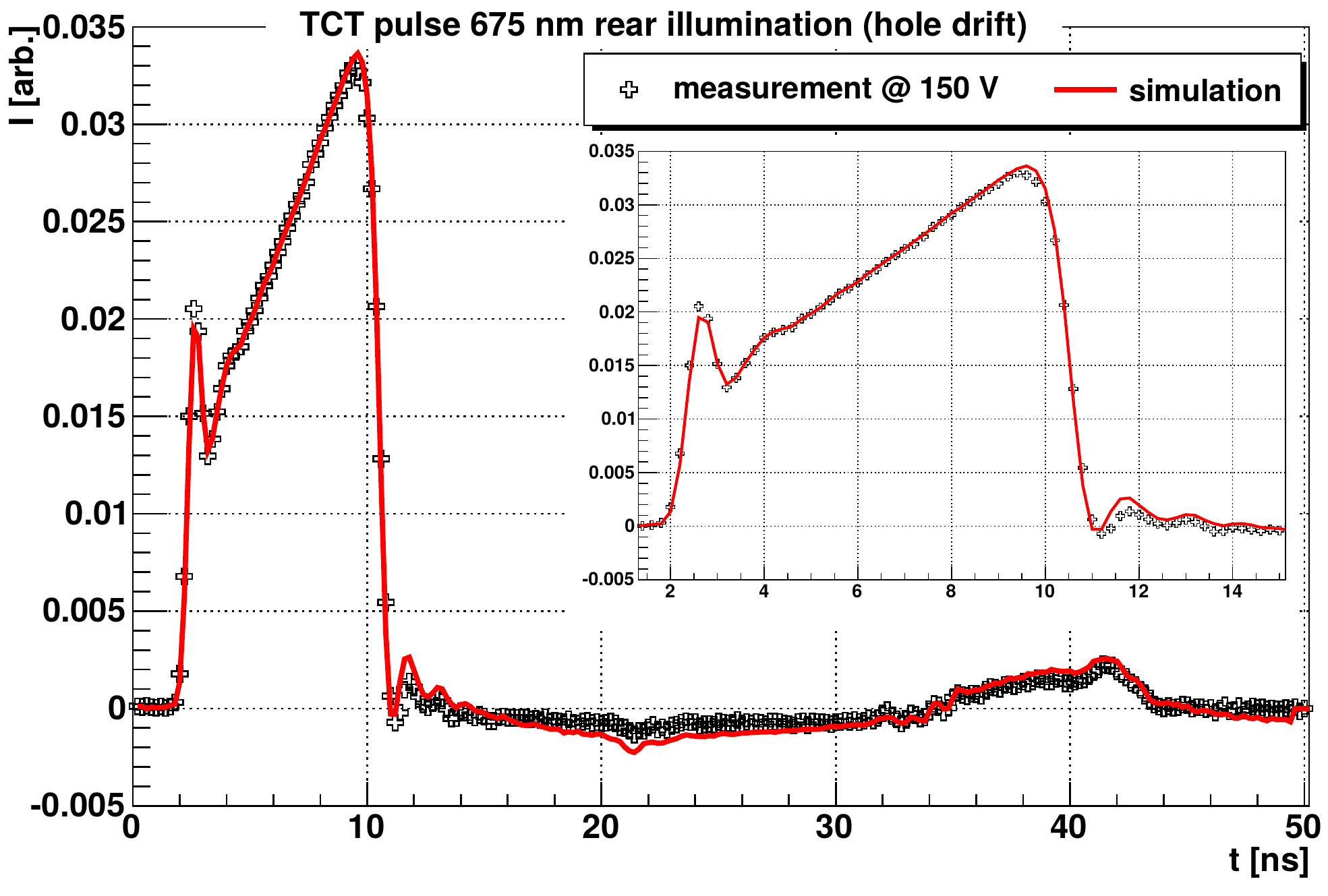}
    \caption{ }
     \label{fig:H150}
   \end{subfigure}%
   \caption{\,Comparison of the measured current transients (crosses) with the simulated ones (solid lines) for the HPK n-type diode using 675~nm laser light measured at $\unit[T=313]{K}$.
    (a) Front-side illumination at $\unit[U_{bias}=1000]{V}$ (electron signal),
    (b) rear-side illumination at $\unit[U_{bias}=1000]{V}$ (hole signal),
    (c) front-side illumination at $\unit[U_{bias}=150]{V}$ (electron signal), and
    (d) rear-side illumination at $\unit[U_{bias}=150]{V}$ (hole signal).}
  \label{fig:Example-fit}
 \end{figure}

 As shown in more detail in Ref.~\citep{transferf}, the entire data set is well described by the fits.
 To illustrate the quality of the description of the data, we show four examples in Fig.~\ref{fig:Example-fit}:
 the transients for $ U_{bias} = 1000$ and $\unit[150]{V}$ at $\unit[T=313]{K}$ for the front- and  the rear-side illumination with the 675~nm laser light.
 Details of the pulse shape, including signal reflections at connectors and the amplifier, are well described.

 \subsection{Time-of-flight method}
  \label{subsect:Tof}

 The well established time-of-flight (tof) method, used e.g. in Ref.~\citep{canali1971drift}, was used as a cross-check of the results of the fit by the simulated transients discussed above.
 As shown below, the tof method can be extended to the situation of non-uniform electric fields, if the inverse mobility is a linear function of the electric field, as it is the case for the parametrization of Eq.\,\ref{eq:newparlin} below and above the field $E_0$.
 The data with front-side injection of the 675~nm laser were used for determining the electron drift velocity, and the ones with rear-side 675~nm light injection for the hole drift velocity.
 For the time-of-flight, $t_{tof}$, the time difference between the maximum slopes at the rise and at the fall of the pulse has been used.
 Using the simulation discussed above, we have verified, that this definition of $t_{tof}$ agrees with the calculated transit time $\int_0^w \big(\mu(E) \cdot E(x)\big)^{-1}\mathrm{d}x$ to within 100~ps for $\unit[U_{bias} > 150]{V}$.

 As the electric field, $E(x)$, depends on the position in the sensor, $x$, also the drift velocity $v_d(x)$ depends on $x$. The electric field, $E_{tof}$, corresponding to the drift velocity $v_d(E_{tof}) = w/t_{tof}$ obtained by the time-of-flight measurement, has to be determined.
 As shown below, $E_{tof} = \langle 1/E(x) \rangle ^{-1}$,  with $\langle \, \rangle$ denoting the average over $x$, if the inverse of the mobility is constant or linear in $E$.
 These conditions are satisfied if the mobility can be described by Eq.~\ref{eq:newparlin} and the electric field in the entire sensor is either smaller or larger than $E_0$.

 Here the argument: from $1/\mu (E) = a_1 + a_2 \cdot E(x)$
 follows $1/v_d(E) = a_1 /E(x) + a_2$, and from
 $t_{tof} = \int_0^w(1/v_d(E))\,\mathrm{d}x =
 \int_0^w(a_1 /E(x) + a_2)\,\mathrm{d}x = w \cdot (a_1 \,\langle 1/E(x) \rangle + a_2)$
 and $t_{tof} = w/v_d(E_{tof})$, follows $1/E_{tof} = 1/\langle E(x) \rangle$.
 For uniform doping $E_{tof} = (E_{max}-E_{min})/\ln(E_{max}/E_{min})$, where $E_{max}$ and $E_{min}$ denote the maximum and minimal fields in the sensor, respectively.

 For a sensor of 200\,$\upmu $m thickness and 90\,V depletion voltage, for bias voltages of 100, 150, 250, 500, and $\unit[1000]{V}$ the values for $\langle E \rangle$ are: 5, 7.5, 12.5, 25, and $\unit[50]{kV/cm}$. The corresponding values for $\langle 1/E \rangle ^{-1}$ are 3.06, 6.49, 11.9, 24.7, and $\unit[49.9]{kV/cm}$.
 The differences between $\langle E \rangle$ and $\langle 1/E \rangle ^{-1}$, in particular for bias voltages close to the depletion voltage, are significant.

 As $1/v_{d}(E)= a_1/E(x) + a_2$ is only an approximation for Eq.~\ref{eq:munewh}, we  also compared the measured values of $t_{tof}$ with the values of $t_{tof} = \int_0^w \big(\mu(E) \cdot E(x)\big)^{-1}\mathrm{d}x$ using the values for $\mu (E)$ from the the parametrization Eq.~\ref{eq:newparlin} for electrons and Eq.~\ref{eq:munewh} for holes.
 However, the differences to the method described above are small, and the linear dependence of the inverse mobility provides an adequate approximation for the tof method even close to the depletion voltage, where the variations of $E(x)$ and $v_d(x)$ are large.

\section{Results}
 \label{sect:Results}

  \subsection{Fit and time-of-flight results}
  \label{subsect:FitResults}

 We first fitted separately the data of the HPK n-type diodes for the individual temperatures using the parametrization of Eq.\,\ref{eq:newparlin} for electrons and of Eq.\,\ref{eq:munewh} for holes.
 As the minimum investigated mean electric field was $\unit[5]{kV/cm}$, our data are not sensitive to the constant low-field mobility of electrons below $\unit[1.8]{kV/cm}$ at room temperature. For reaching lower field values thicker sensors with higher resistivity as used by Canali et al. are required.
 The results are given in Table\,\ref{tab:T_Fit}.

  \begin{table}  [!ht]
   \begin{tabular}{|c||c|c||c|c|c|c|}
    \cline{2-7}
    \multicolumn{1}{c|}{} & \multicolumn{2}{c||}{Electrons} & \multicolumn{4}{c|}{Holes}\tabularnewline
   \hline
     $\unit[T]{[K]}$ & $\unit[\mu_{0}]{[cm^{2}/Vs]}$ & $\unit[v_{sat}]{[10^{7}cm/s]}$ & $\unit[\mu_{0}]{[cm^{2}/Vs]}$ & $\unit[b]{[10^{-7}s/cm]}$ & $\unit[c]{[10^{-13}s/V]}$ & $\unit[E_{0}]{[V/cm]}$\tabularnewline
  \hline
  \hline
   233 & 2333 & 1.127 & 842.6 & 1.040 & -4.899 & 2096\tabularnewline
  \hline
   253 & 2013 & 1.103 & 697.9 & 1.024 & -4.608 & 2123\tabularnewline
  \hline
   263 & 1882 & 1.089 & 631.9 & 1.019 & -4.558 & 2379 \tabularnewline
  \hline
   273 & 1753 & 1.078 & 576.8 & 1.003 & -4.293 & 2476\tabularnewline
  \hline
   293 & 1523 & 1.055 & 486.1 & 0.9705 & -3.728 & 2667 \tabularnewline
  \hline
   313 & 1327 & 1.031 & 408.8 & 0.9390 & -3.229 & 3313\tabularnewline
  \hline
   333 & 1157 & 1.014 & 349.2 & 0.8938 & -2.443 & 3980\tabularnewline
  \hline
 \end{tabular}
  \caption{Results of the fit of the simulated to the measured transients at the individual temperatures $T$ for the HPK n-type sensor, using Eq.\,\ref{eq:newparlin} for the parametrization of the field dependence of the mobility of electrons, and Eq.\,\ref{eq:munewh} for the one of holes.
 \label{tab:T_Fit}}
\end{table}

 We then performed a global fit to the entire data set: the HPK n-type sensor illuminated with red laser light from the front and rear side as well as with the infrared laser, for temperatures between 233 and $\unit[333]{K}$ and voltages between 100 and $\unit[1000]{V}$ in steps of $\unit[10]{V}$.
 The corresponding mean electric fields are between 5 and $\unit[50]{kV/cm}$.
 Overall, more than $10^{5}$ data points, each one consisting of 512 measurements, were used for the fit.
 For every iteration the transient for every data set had to be  simulated.
 For the temperature dependence of the parameters, the parametrization of Eq.\,\ref{eq:parT} has been used.
 The results are given in Table~\ref{tab:Temp-dep-par}.

\begin{figure}[!ht]
  \centering
   \begin{subfigure}[a]{7.5cm}
    \includegraphics[width=7.5cm]{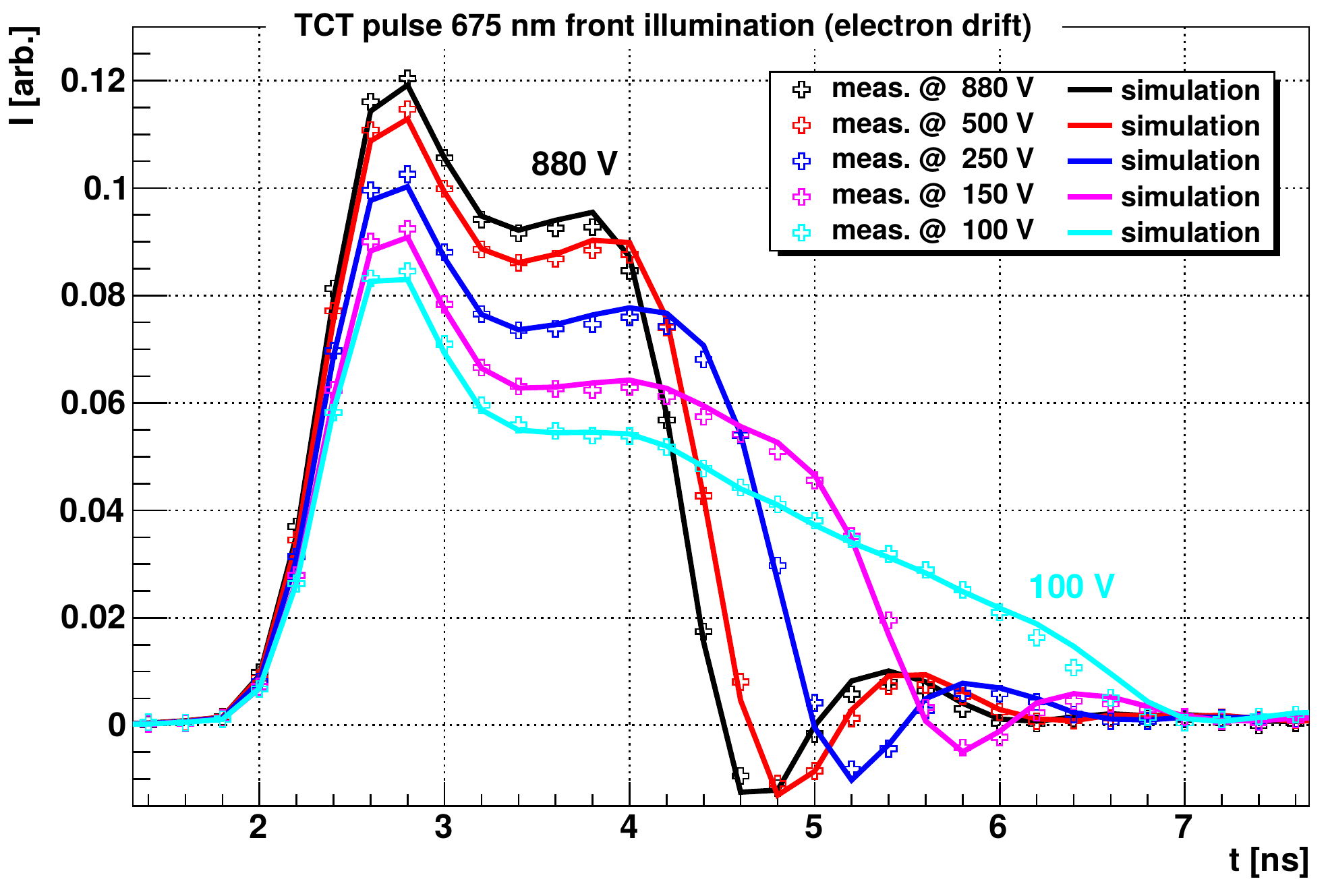}
    \caption{ }
     \label{fig:E233K}
   \end{subfigure}%
    ~
   \begin{subfigure}[a]{7.5cm}
    \includegraphics[width=7.5cm]{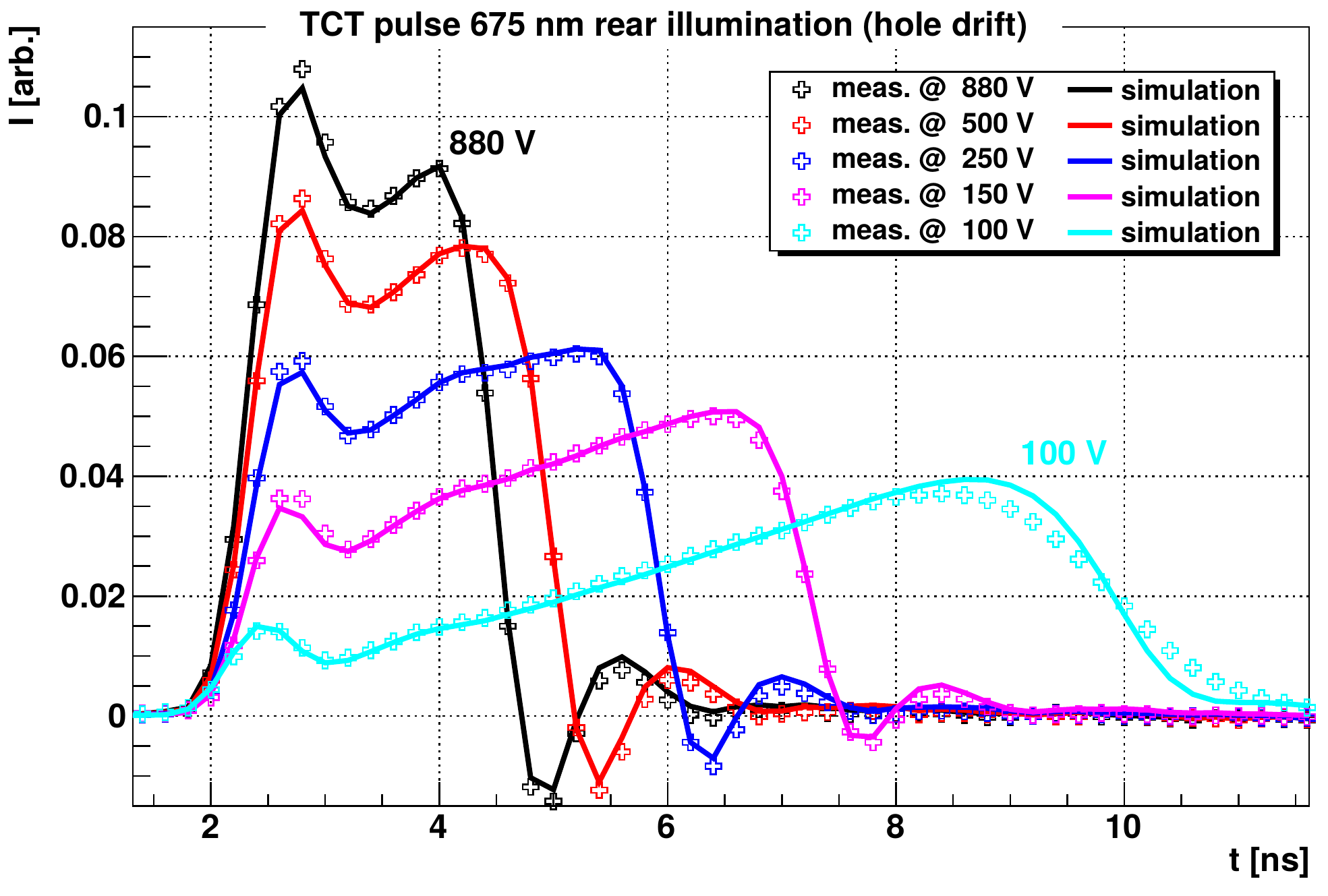}
    \caption{ }
     \label{fig:H233K}
   \end{subfigure}
    ~
   \begin{subfigure}[a]{7.5cm}
    \includegraphics[width=7.5cm]{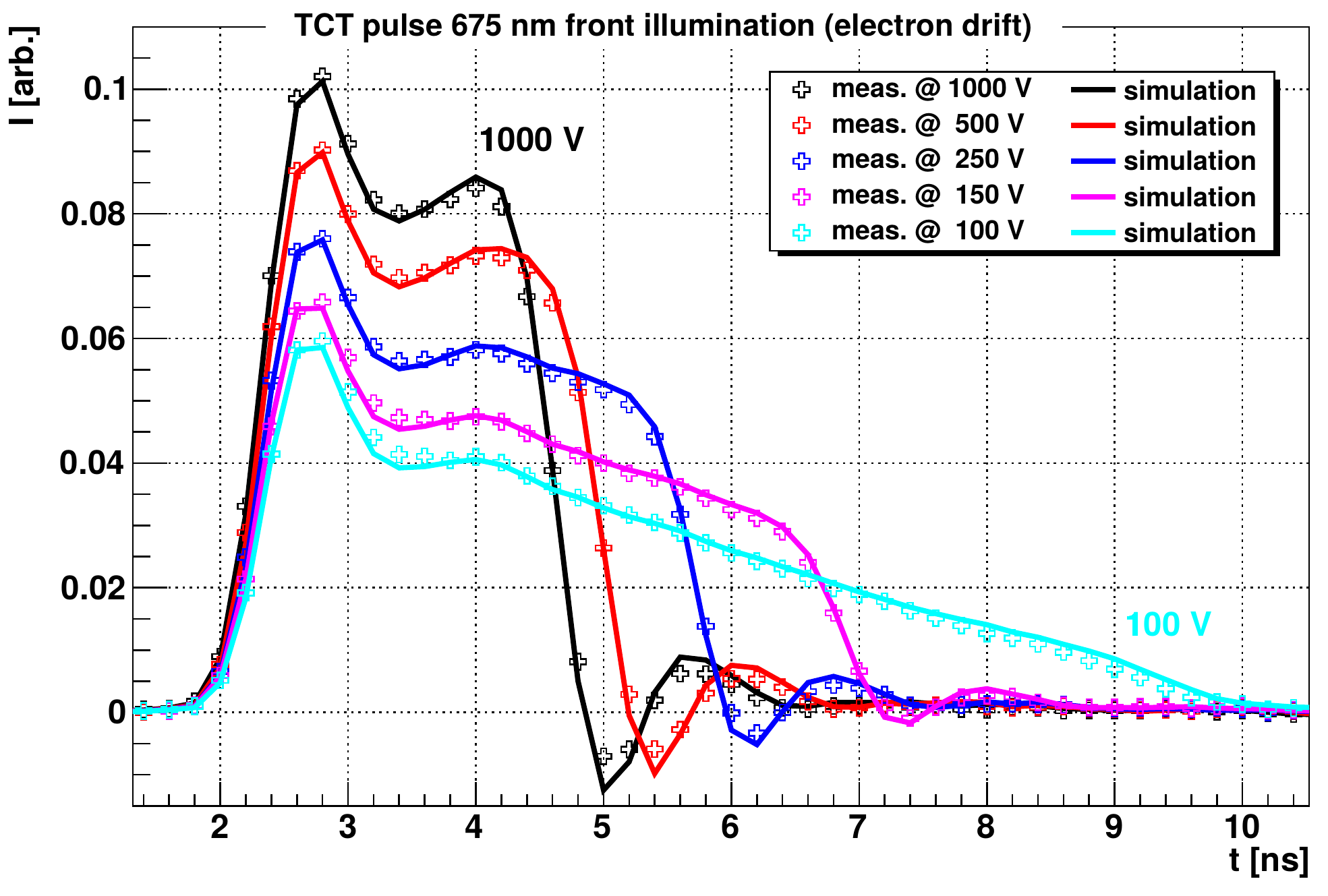}
    \caption{ }
     \label{fig:E333K}
   \end{subfigure}%
    ~
   \begin{subfigure}[a]{7.5cm}
    \includegraphics[width=7.5cm]{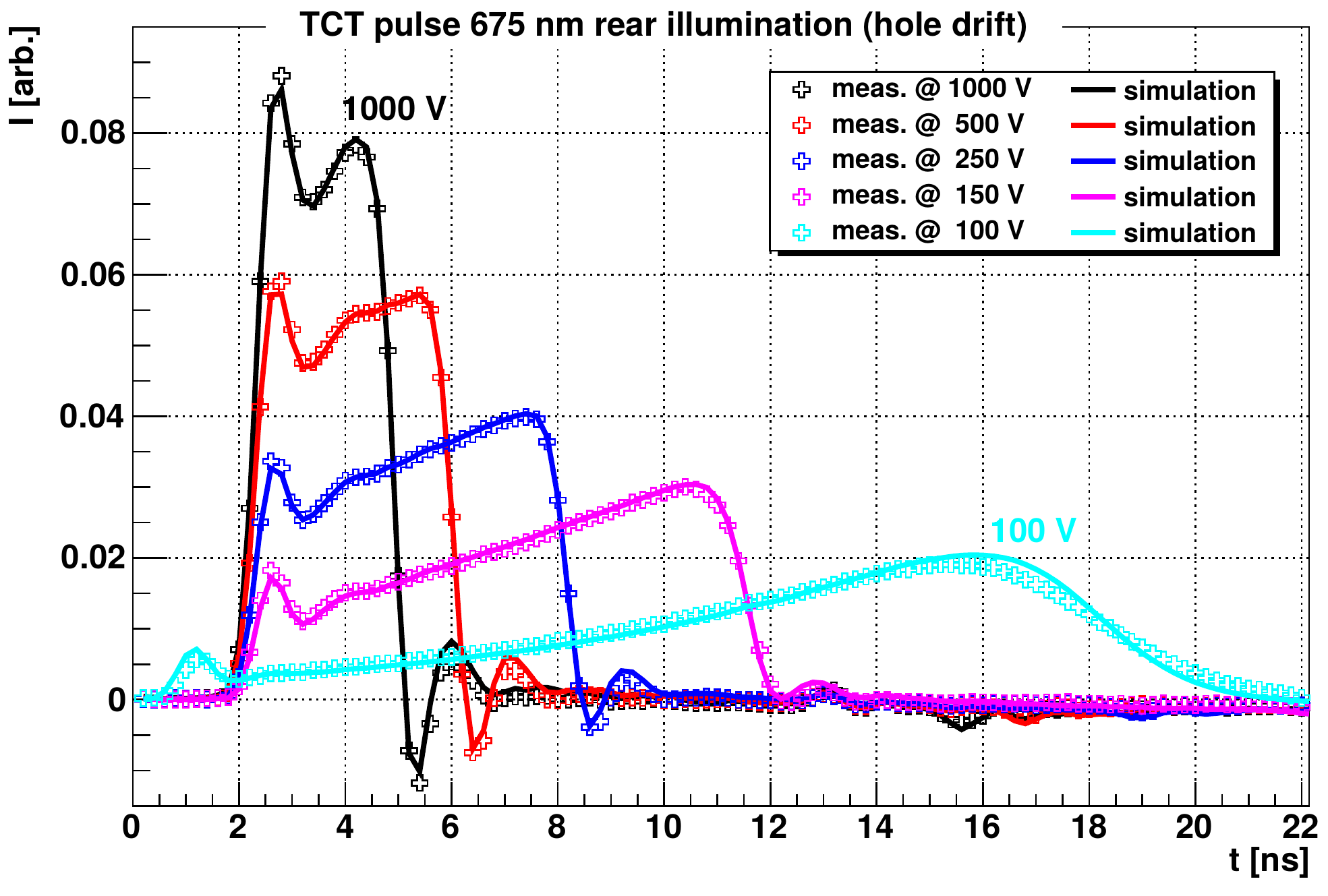}
    \caption{ }
     \label{fig:H333K}
   \end{subfigure}%
   \caption{\,Comparison of the measured current transients (crosses) with the simulated ones (solid lines) for the global fit discussed in the text for the HPK n-type diode using 675~nm laser light and bias voltages between 100 and $\unit[1000]{V}$.
    (a) Front-side illumination at $\unit[233]{K}$ (electron signal),
    (b) rear-side illumination at $\unit[233]{K}$ (hole signal),
    (c) front-side illumination at $\unit[333]{K}$ (electron signal), and
    (d) rear-side illumination at $\unit[333]{K}$ (hole signal).}
  \label{fig:Global-fit}
 \end{figure}

 The entire data set is well described by the global fit.
 To illustrate the quality of the description, Fig.\,\ref{fig:Global-fit} shows the transients for different bias voltages for the lowest ($\unit[233]{K}$)  and highest ($\unit[333]{K}$) temperatures for front- and rear-side-illumination with the 675\,nm laser light.
 The biggest differences between fit and data are observed in Fig.\,\ref{fig:H233K} around 10\,ns at a bias voltage of $\unit[100]{V}$.
 As this voltage is only about $\unit[10]{V}$ above the depletion voltage, the charge carriers are generated in a low field region, where the plasma effect\,\citep{Tove1967, Becker2010Plasma} may cause an additional delay and diffusion for the holes.

\begin{table}
\begin{centering}
\begin{tabular}{|c||c|c||c|}
\cline{3-4}
\multicolumn{1}{c}{} &  & $par_{i}(\unit[T=300]{K})$ & $\alpha_{i}$\tabularnewline
\hline
Electrons & $\mu_{0}^{e}$ & $\unit[1430]{cm^{2}/Vs}$ & $-1.99$\tabularnewline
\cline{2-4}
 & $v_{sat}^{e}$ & $\unit[1.05\cdot10^{7}]{cm/s}$ & $-0.302$\tabularnewline
\hline
\hline
Holes & $\mu_{0}^{h}$ & $\unit[457]{cm^{2}/Vs}$ & $-2.80$\tabularnewline
\cline{2-4}
 & $b$ & $\unit[9.57\cdot10^{-8}]{s/cm}$ & $-0.155$\tabularnewline
\cline{2-4}
 & $c$ & $\unit[-3.24\cdot10^{-13}]{s/V}$ & $-$\tabularnewline
\cline{2-4}
 & $E_{0}$ & $\unit[2970]{V/cm}$ & $5.63$\tabularnewline
\hline
\end{tabular}
\par\end{centering}
\centering{}
 \caption{Parameters for the mobility for <100> silicon obtained from the fit to the data of the HPK n-type diode using Eq.~\ref{eq:munewe} for the electron mobility, Eq.~\ref{eq:munewh} for the hole mobility, and Eq.~\ref{eq:parT} for the temperature dependence. The mean electric field values range from 5 to $\unit[50]{kV/cm}$, and the temperatures from 233 to $\unit[333]{K}$. The uncertainty is estimated to be $\unit[2.5]{\%}$ for the electron and $\unit[5]{\%}$ for the hole mobility.
 \label{tab:Temp-dep-par}}
\end{table}

\begin{figure}[!ht]
  \centering
   \begin{subfigure}[a]{7.5cm}
    \includegraphics[width=7.5cm]{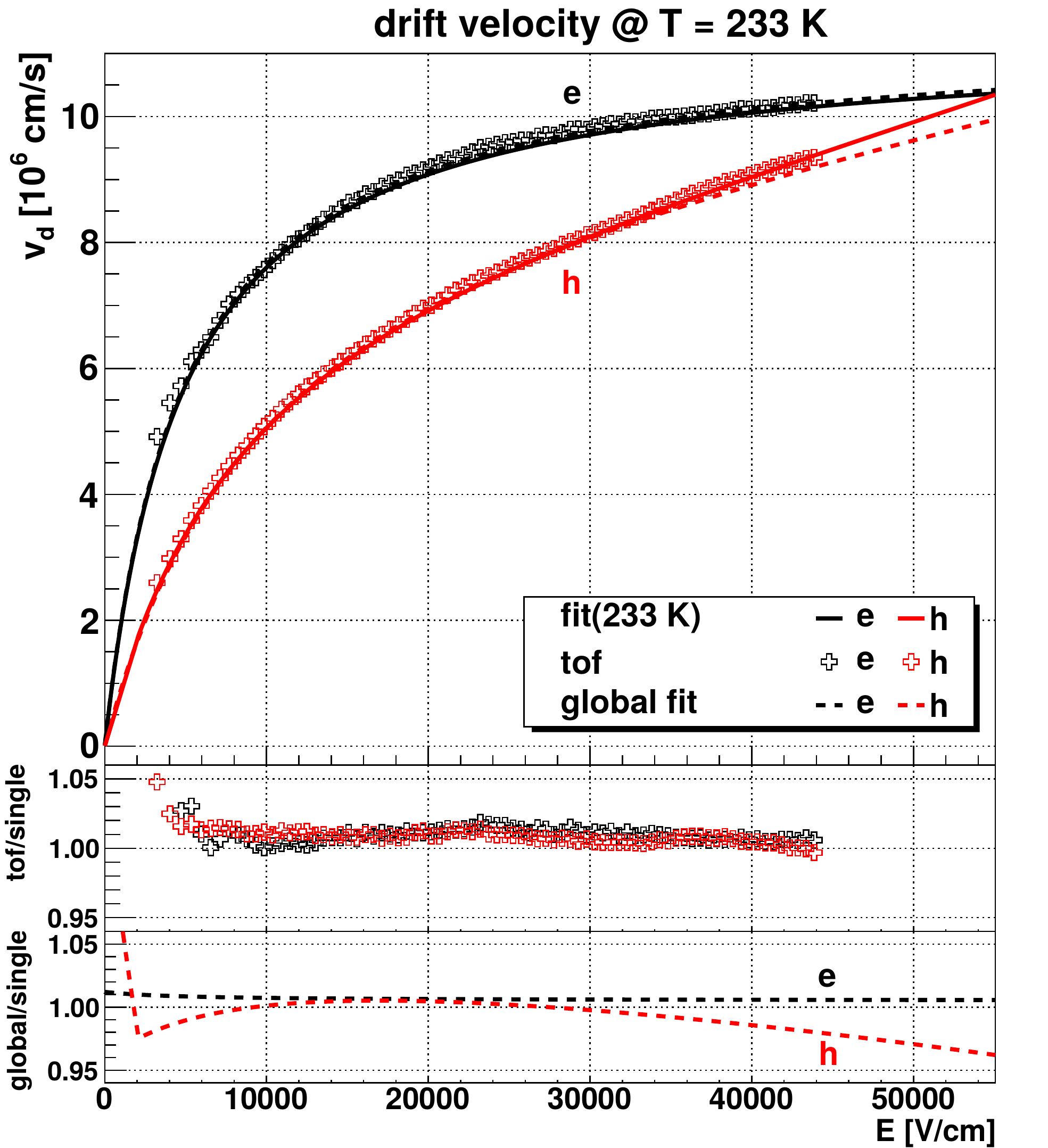}
    \caption{ }
     \label{fig:233K_stg}
   \end{subfigure}%
    ~
   \begin{subfigure}[a]{7.5cm}
    \includegraphics[width=7.5cm]{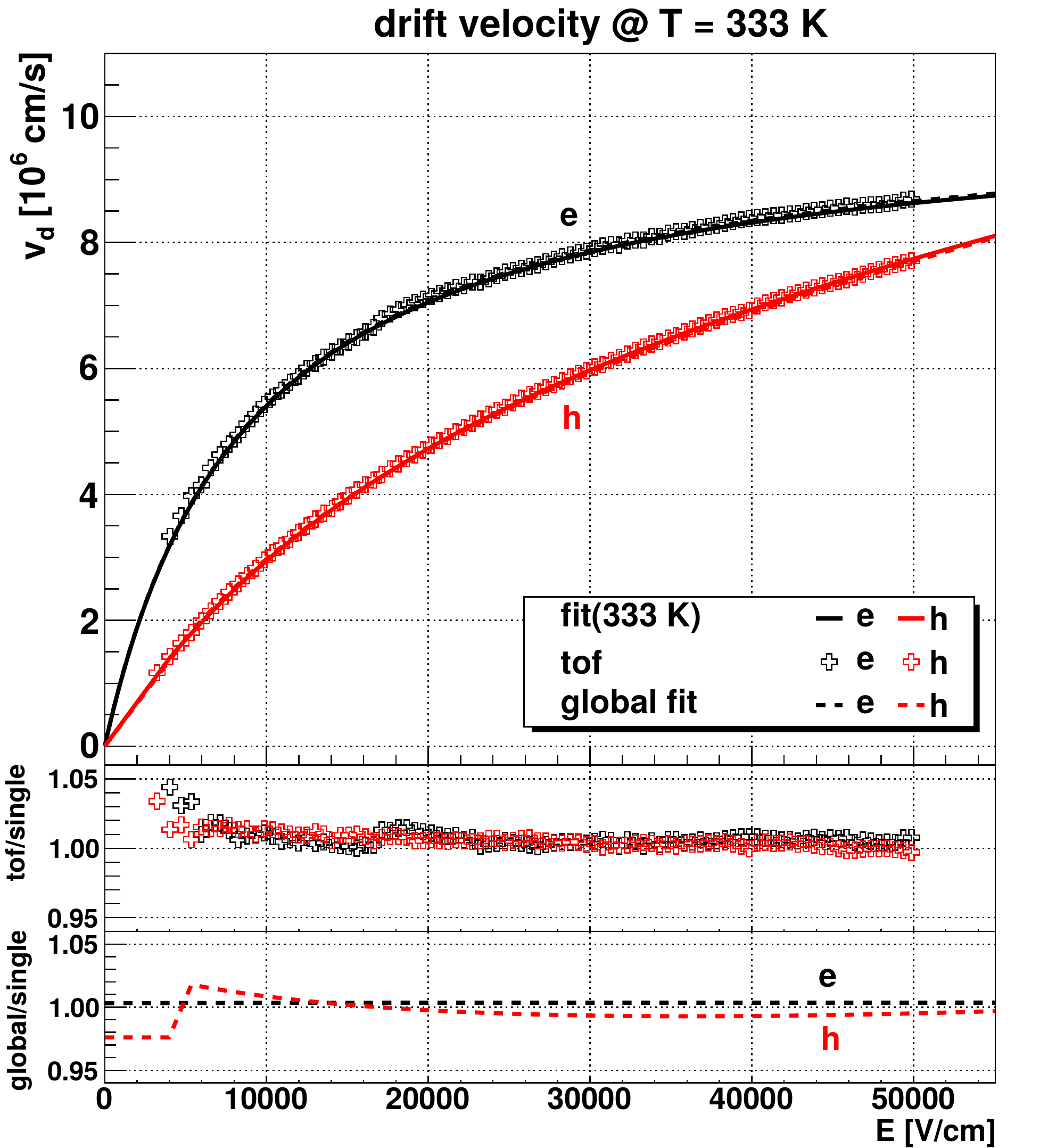}
    \caption{ }
     \label{fig:333K_stg}
   \end{subfigure}%
   \caption{\,Top: Comparison of the drift velocities of electrons (black) and holes (red) from the tof measurements (crosses), the fits at a single temperature (solid lines), and the global fit (dashed lines).
   Middle: Ratio of the tof to the single-temperature-fit results, and bottom: Ratio of the global-fit to the single-temperature-fit results.
    (a) Results for $\unit[233]{K}$, and
    (b) for $\unit[333]{K}$.}
  \label{fig:T_stg}
 \end{figure}

 In Fig.\,\ref{fig:T_stg} we compare for 233 and $\unit[333]{K}$ the field dependencies of the electron and hole velocities from the fit at this particular temperature to the ones from the global fit and from the tof measurements.
 The drift velocities are shown on top, the ratio of the velocities from the tof measurement to the fit at the particular temperature in the middle, and the ratio of the velocities from the global fit to the fit at the particular temperature at the bottom.
 We note that, with the exception of electric fields below 5\,kV/cm where the tof measurements become inaccurate, the results from the tof measurement and the fit at a single temperature agree within about 1\,\%.
 This gives an idea of the accuracy of the analysis methods.
 The comparison of the global-fit results to the single-temperature-fit results shows that, for the electric fields where the measurements were made, the agreement is within about 2\,\%.
 However, extrapolating to lower and higher field values, the differences, in particular for holes at $T = \unit[233]{K}$, rapidly increase.

 \subsection{Comparison of different diodes}
  \label{subsect:ComparisonDiodes}

 In order to check the results from the HPK n-type diode, we also performed measurements on a $\unit[290]{\upmu m}$-thick n-type diode produced by CiS and a $\unit[200]{\upmu m}$-thick p-type diode produced by HPK, with the parameters shown in Table\,\ref{tab:Properties-samples}.
 We expect different systematic errors for the different diodes.
 As an example, the full-width-at-half-maximum of the transfer function is about $\unit[0.6]{ns}$ for the HPK diodes, and about $\unit[1.0]{ns}$ for the CiS diode, which has a 5.6 times larger area and is $\unit[40]{\%}$ thicker.

\begin{figure}
  \centering
   \begin{subfigure}[a]{7.5cm}
    \includegraphics[width=7.5cm]{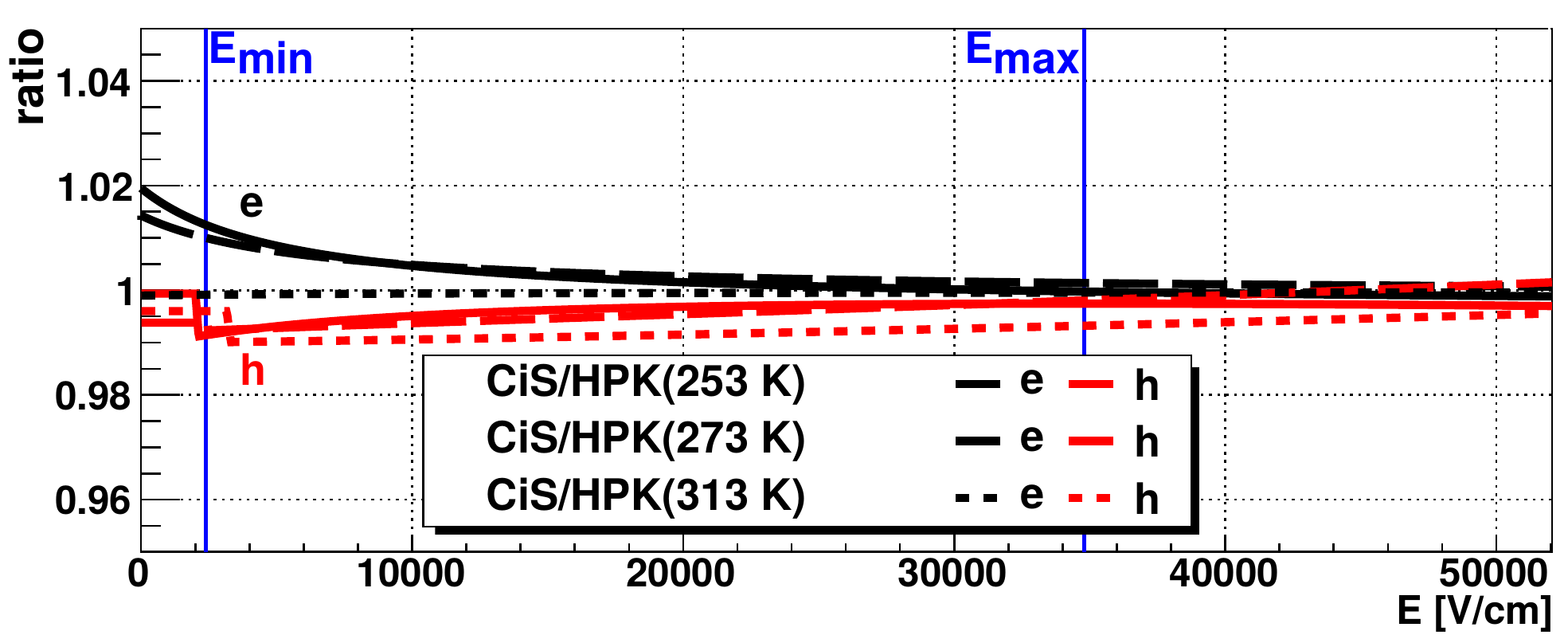}
    \caption{ }
     \label{fig:vgl_CIS_rev}
   \end{subfigure}%
    ~
   \begin{subfigure}[a]{7.5cm}
    \includegraphics[width=7.5cm]{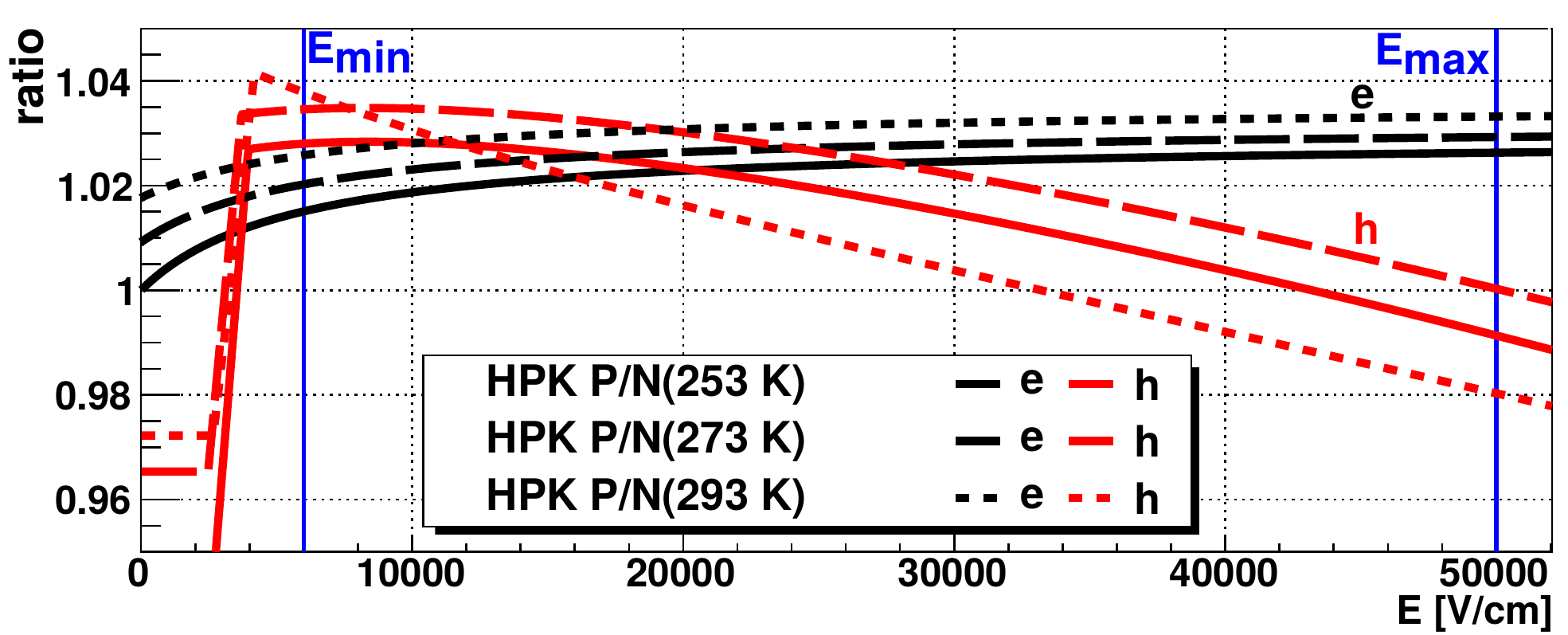}
    \caption{ }
     \label{fig:vgl_pHPK_rev}
    \end{subfigure}%
   \caption{
   Ratios of the measured drift velocities of electrons and holes at temperatures of 253, 273 and $\unit[293/313]{K}$ for
   (a)\,CIS n-type diode to HPK n-type diode, and
   (b)\,HPK p-type diode to HPK n-type diode.
   The vertical lines denote the maximum and minimum mean electric fields in the diodes.}
  \label{fig:vgl}
 \end{figure}

 As an illustration we show the comparisons of the drift velocities for electrons and holes as function of the electric field at 253, 273 and $\unit[293/313]{K}$.
 Fig.\,\ref{fig:vgl_CIS_rev} shows the ratios of the drift velocities for the CIS n-type diode relative to the HPK n-type diode.
 For both electrons and holes the results agree within $\unit[1]{\%}$ for fields above $\unit[2.5]{kV/cm}$, where our measurements are sensitive.
 Fig.\,\ref{fig:vgl_pHPK_rev} shows the comparison of the HPK p-type to the HPK n-type diode.
 The electron velocity is systematically higher by about $\unit[2-3]{\%}$ and the hole velocity higher by about $\unit[3]{\%}$ at a voltage of $\unit[6]{kV/cm}$, and lower by about $\unit[1]{\%}$ at $\unit[50]{kV/cm}$.
 The measurements are not sensitive to fields below $\unit[2.5]{kV/cm}$, where the differences are larger.
 We conclude that over the entire electric field range of our measurements, the drift velocities agree to better than  $\unit[4]{\%}$.

 We note that we did not achieve a good description of the transients of the HPK p-type diode with the electric field used so far: a linear function from the diode surface at the pn-junction side to the other surface.
 However, a good description was achieved by introducing field-free regions of 1 to 2\,$\upmu $m at the diode surfaces.
 In the simulation charge carriers, which are produced in the field-free regions, reach the electrodes or the field region by diffusion and thus change the transients.
 For more details we refer to Appendix\,2.
 We also note that the introduction of the field-free regions hardly affects the values obtained for the drift velocities from the fits.
 
 From the differences between the two methods and between the fit results, we estimate for fields between 2.5 and $\unit[50]{kV/cm}$ an uncertainty of $\unit[2.5]{\%}$ for the electron and of $\unit[5]{\%}$ for hole drift velocities for the values of the parameters given in table~\ref{tab:Temp-dep-par}.

 \subsection{Comparison with published data}
  \label{subsect:ComparisonLiterature}

\begin{figure}
  \centering
   \begin{subfigure}[a]{7.5cm}
    \includegraphics[width=7.5cm]{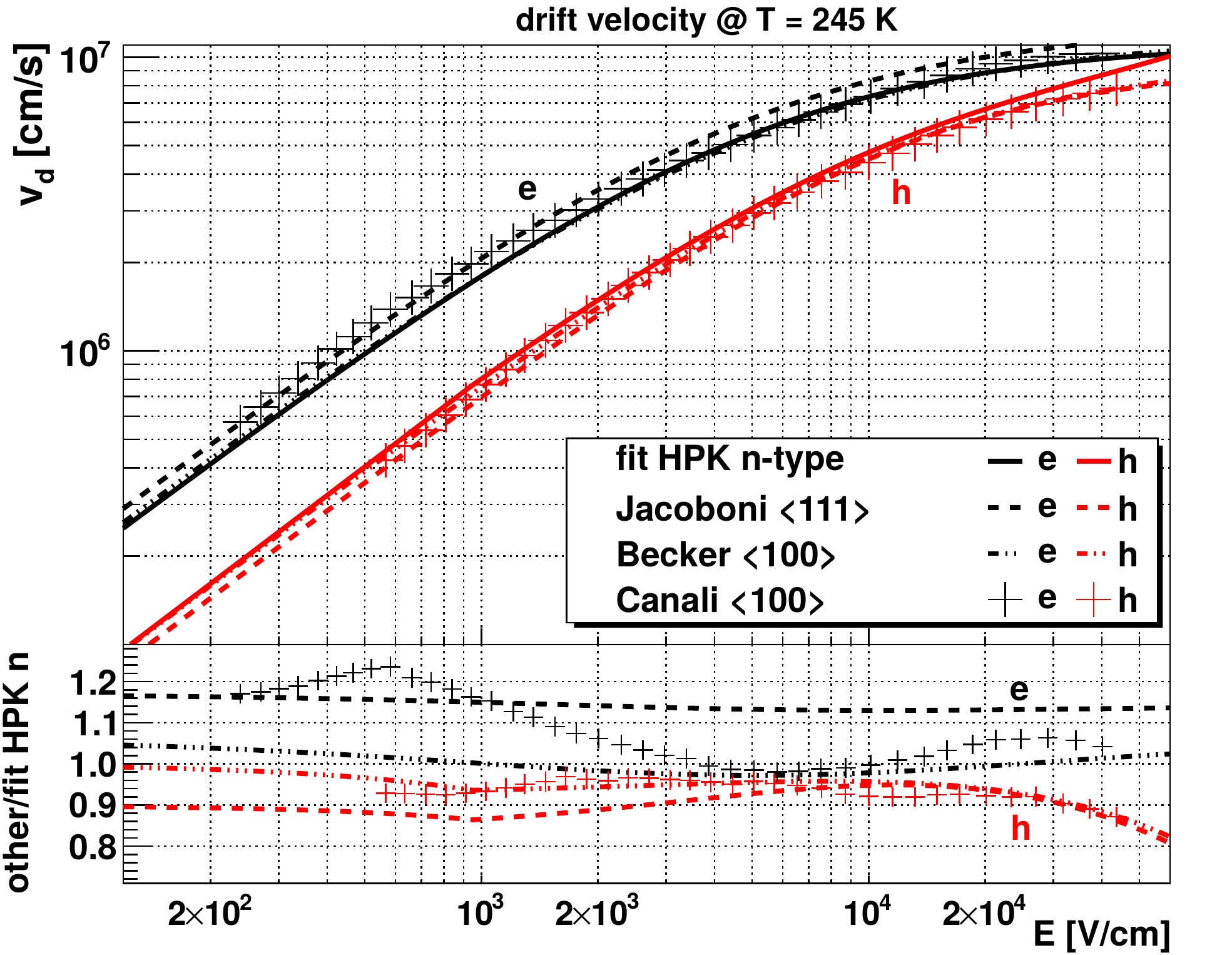}
    \caption{ }
     \label{fig:245K_vgllit}
   \end{subfigure}%
    ~
   \begin{subfigure}[a]{7.5cm}
    \includegraphics[width=7.5cm]{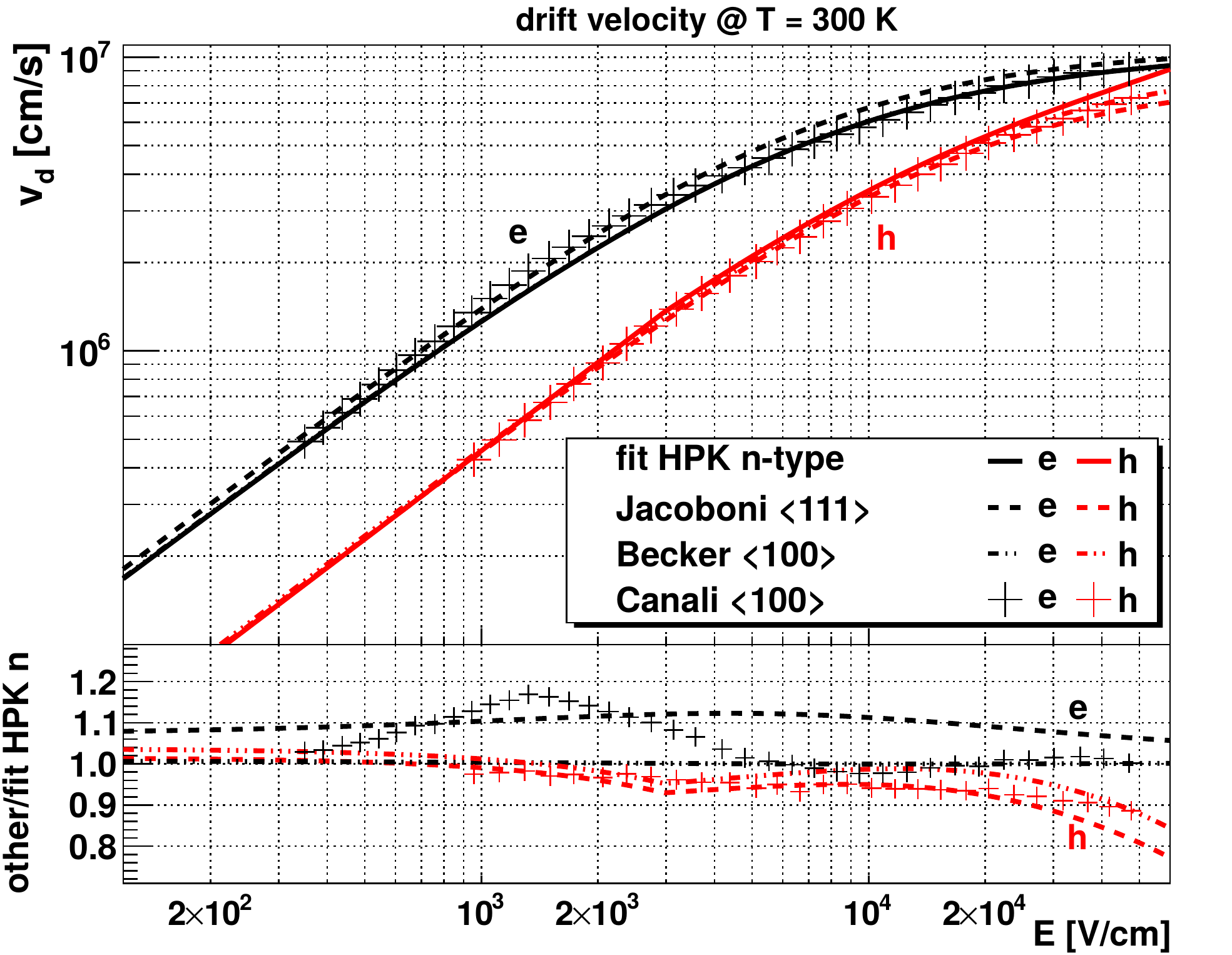}
    \caption{ }
     \label{fig:300K_vgllit}
    \end{subfigure}%
   \caption{
   Comparison of the drift velocities from the global fit results of data from the HPK n-type diode to published data.
   On the top the drift velocities and on the bottom their ratios are shown for
   (a) $\unit[245]{K}$, and
   (b) $\unit[300]{K}$.   }
  \label{fig:Comp-fit-lit}
 \end{figure}

 We compare our results for <100> silicon, presented in Table~\ref{tab:Temp-dep-par}, to the results of Jacoboni et al.~\citep{jacoboni1977review} for <111>, of Becker et al.~\citep{becker2011measurements} for  <100>, and the TOF results of Canali et al.~\citep{canali1971drift} for <100> silicon.
 We also compare to the Jacoboni <111> parametrization~\cite{jacoboni1977review}, because it is widely used for <100> silicon due to the lack of publications on <100> drift velocities.
 Figure~\ref{fig:Comp-fit-lit} shows our results and the published values at 245 and $\unit[300]{K}$.
 On top the drift velocities are shown, and at the bottom the ratios to our global-fit results from the HPK n-type diode.

 At both temperatures the Jacoboni <111> drift velocities for electrons are $10 - 20$\,\% higher than our <100> results.
 At $\unit[245]{K}$ the Jacoboni <111> drift velocity for holes is $2 - 10$\,\% lower than our <100> results.
 At $\unit[300]{K}$ the differences, in particular for electric fields below 10\,kV/cm, are smaller.

 When comparing our results to the values published by Becker et al., one notices that the electron drift velocities are very similar, whilst the drift velocities for holes are $\sim \unit[15]{\%}$ below our results at high electric fields.
 However, the measurements of Becker et al. were made at $\langle E \rangle = (\unit[3.8,\,7.1,\,17.9)]{kV/cm}$, where the differences are less than $\unit[6]{\%}$, and the large differences are outside of the region of their measurements.

 When comparing the Canali et al. tof values to our fit results, one notices that the electron drift velocities are up to $\unit[20 - 25]{\%}$ higher for fields below $\unit[2]{kV/cm}$. However, these field values are outside of the range of our measurements.
 The hole drift velocities determined by Canali et al. are similar to our results at low electric fields, but at high electric fields they are up to $\unit[15]{\%}$ lower.

 \subsection{Parametrization of the drift velocities for fields up to 50~kV/cm}
 In order to provide a parametrization of the drift velocities of holes and electrons in <100> silicon, which can be directly used in simulations, we have combined the low-field data for <111> silicon of Ref.~\cite{jacoboni1977review} with our data. The low-field mobility in silicon is isotropic as shown in Ref.~\cite{jacoboni1977review}.
 For electrons the parametrization of Eq.~\ref{eq:munewe}, for holes the one of Eq.~\ref{eq:munewh}, and for the temperature dependencies the one of Eq.~\ref{eq:parT} have been used. For the fit the values of $\mu_0$ and $\alpha_i$ of Ref.~\cite{jacoboni1977review} were fixed. 
 The results are given in Table~\ref{tab:Temp-all-fields}.
 
 We want to stress that we did not observe the expected constant part of the electron mobility. The transition region of the electron mobility from a constant value to a linear increase, corresponding to velocity saturation, can not be described by the simple model presented in Eq.~\ref{eq:newparlin}. But, at least one additional parameter similar to the quadratic term $c$ for the hole mobility, Eq.~\ref{eq:munewh}, is needed for an adequate description. Unfortunately, our data in the transition region between about 0.5 and $\unit[2.5]{kV/cm}$ is not sufficient to determine such a parameter precisely.
 
 For fields between 2.5 and $\unit[5]{kV/cm}$ differences of up to $\unit[9]{\%}$ to the fit without fixing the low-field mobility are found. There is no error estimation given for the values of Ref.~\cite{jacoboni1977review}. However, the authors of Ref.~\cite{jacoboni1977review} note the error to be comparable to $\unit[5]{\%}$.  Therefore, we recommend using the parameters of Table~\ref{tab:Temp-dep-par} for $E\geq\unit[2.5]{kV/cm}$.

 \begin{table}
  \begin{centering}
   \begin{tabular}{|c||c|c||c|}
    \cline{3-4}
    \multicolumn{1}{c}{} &  & $par_{i}(\unit[T=300]{K})$ & $\alpha_{i}$\tabularnewline
  \hline
   Electrons & $\mu_{0,Jac}^{e}$ & $\unit[1530]{cm^{2}/Vs}$ & $-2.42$\tabularnewline
  \cline{2-4}
   & $v_{sat}^{e}$ & $\unit[1.03\cdot10^{7}]{cm/s}$ & $-0.226$\tabularnewline
 \hline
 \hline
   Holes & $\mu_{0,Jac}^{h}$ & $\unit[464]{cm^{2}/Vs}$ & $-2.20$\tabularnewline
 \cline{2-4}
   & $b$ & $\unit[9.57\cdot10^{-8}]{s/cm}$ & $-0.101$\tabularnewline
  \cline{2-4}
   & $c$ & $\unit[-3.31\cdot10^{-13}]{s/V}$ & -\tabularnewline
  \cline{2-4}
 & $E_{0}$ & $\unit[2640]{V/cm}$ & $0.526$\tabularnewline
  \hline
 \end{tabular}
\par\end{centering}

\centering{}\protect\caption
 {The mobility parameters for <100> silicon of Eq.~\ref{eq:munewe} for the electron mobility, Eq.~\ref{eq:munewh} for the hole mobility, and Eq.~\ref{eq:parT} for the temperature dependence.
 The values of $\mu_{0,Jac}^{e,h}$ and the corresponding $\alpha_{i}$ values have been taken from~\citep{jacoboni1977review}, and were fixed in the fit to the measurements of this manuscript in order to describe the mobility for electric field values up to
 $\unit[50]{kV/cm}$ and temperatures between 233 and $\unit[333]{K}$. However, for $E\geq\unit[2.5]{kV/cm}$ we recommend using the parameters of Table~\ref{tab:Temp-dep-par}.
 \label{tab:Temp-all-fields}}
\end{table}

\section{Summary}
 \label{sect:Summary}

 The current transients in p- and n-type-silicon pad diodes from electron-hole pairs produced by sub-nanosecond laser light with a wavelength of $\unit[675]{nm}$ injected from both sides, and by laser light of $\unit[1063]{nm}$ injected from the junction side, were measured.
 From these data the drift velocities of electrons and holes in high-ohmic silicon with <100> lattice orientation for electric fields between 2.5 and $\unit[50]{kV/cm}$ and temperatures between 233 and $\unit[333]{K}$ have been determined using two methods:
 a fit of simulated transients
 to the measured transients, and a time-of-flight method.
 The results of both methods agree within $\unit[2]{\%}$.
 New parametrizations of the field dependencies of the drift velocities have been used, which also provide a somewhat better description than previous parametrizations of published results on drift velocities in <111> silicon.
 The measurements were performed in a range of intermediate temperatures which are of actual interest for silicon detectors in high energy physics.

 For electrons our results are similar to published results.
 For holes at high electric fields, significant differences to published results are found.
 The differences to the parametrization for <111> silicon of Jacoboni et al., which is frequently used for the analysis and simulation of <100> silicon, are large.
 By combining published data for the low-field mobility with our data at higher fields, we determine parametrizations of the drift velocities of electrons and holes in <100> silicon for electric fields up to $\unit[50]{kV/cm}$ and temperatures between 233 and $\unit[333]{K}$. Their accuracy is about $\unit[5]{\%}$, and they are available for simulating sensors built on high-ohmic <100> silicon.

\section{Appendices}
 \label{sect:Appendix}

 \subsection{Appendix 1: Comparison of mobility parametrization for <111> silicon}
  \label{subsect:Appendix_Comparison}

\begin{figure}[!ht]
  \centering
   \begin{subfigure}[a]{7.5cm}
    \includegraphics[width=7.5cm]{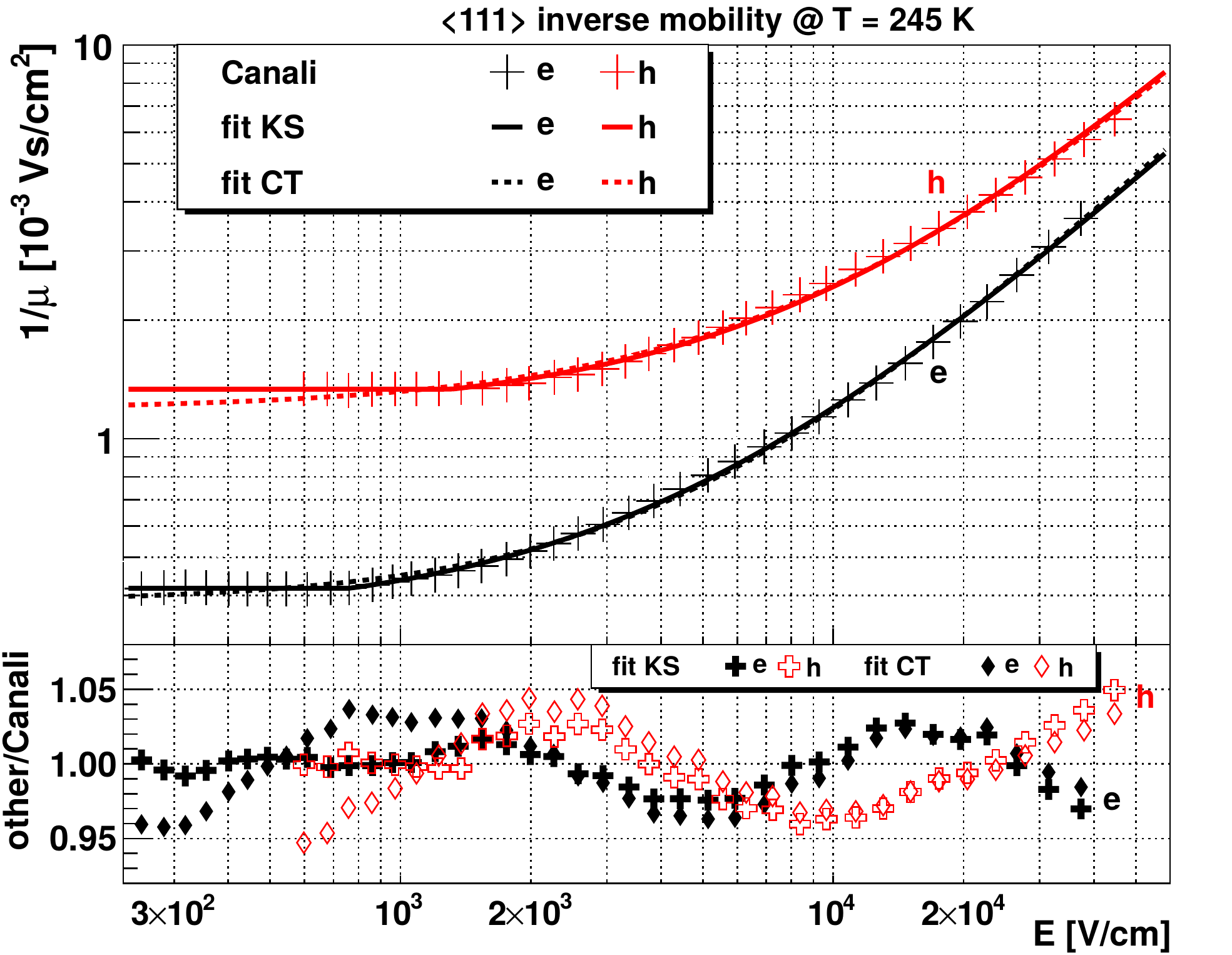}
    \caption{ }
     \label{fig:245K_CTnew}
   \end{subfigure}%
    ~
   \begin{subfigure}[a]{7.5cm}
    \includegraphics[width=7.5cm]{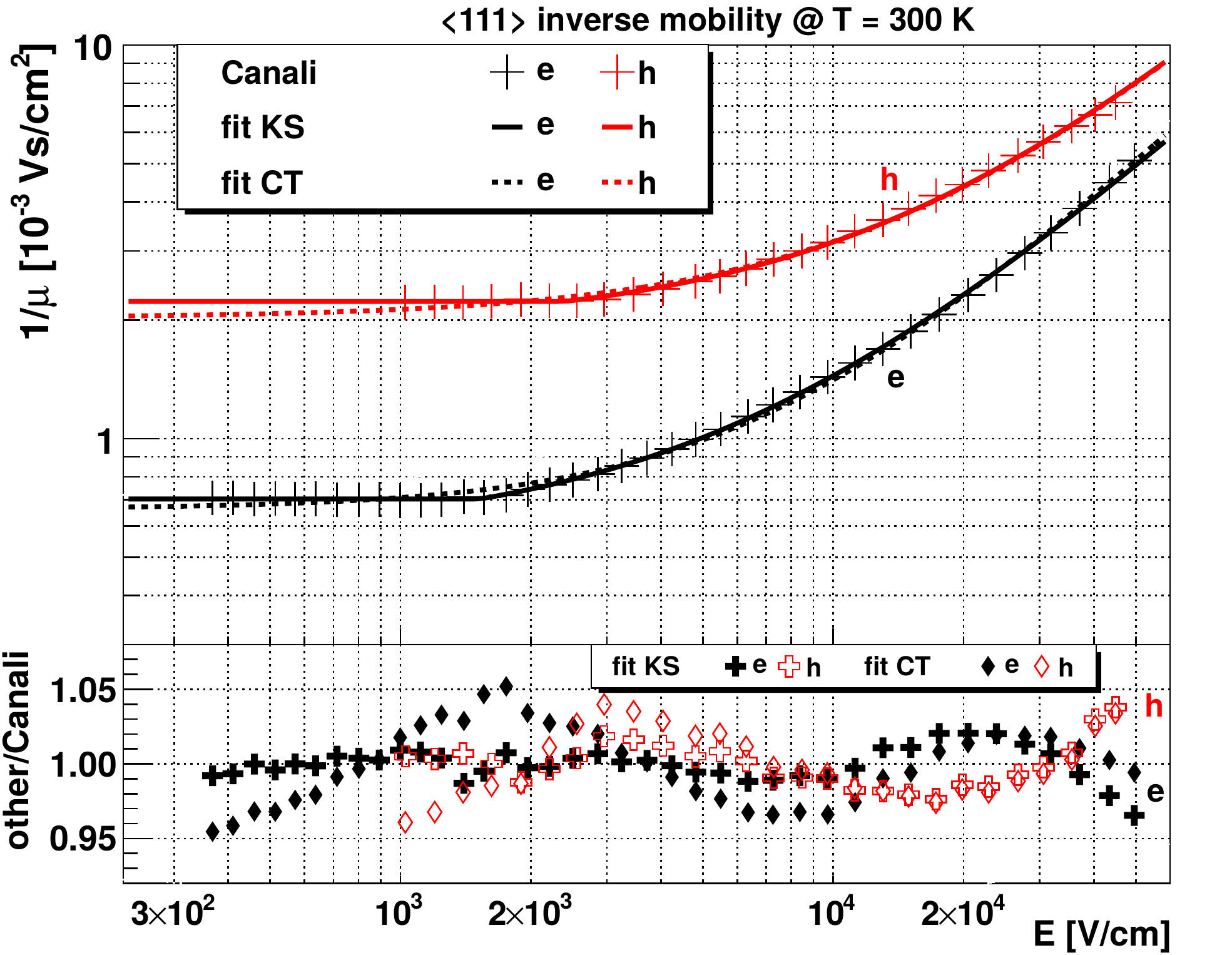}
    \caption{ }
     \label{fig:300K_CTnew}
    \end{subfigure}%
   \caption{
  Inverse mobility, $1/\mu$, of electrons (black) and holes (red) for <111> silicon at (a) $\unit[245]{K}$, and (b) $\unit[300]{K}$.
  The data of Canali~\cite{canali1971drift} (crosses) were fitted by the KS parametrization (Eq.~\ref{eq:newparlin}~$-$~solid lines) and the CT parametrization (Eq.~\ref{CT}~$-$~dashed lines).
  At the bottom the ratio of the Canali data and the values obtained from the fit of the KS (crosses) and the CT parametrization (diamonds) are shown.
   }
  \label{fig:1/MU}
 \end{figure}

 In this appendix we compare fits by the new KS parametrization (Eq.~\ref{eq:newparlin}) and the frequently used CT parametrization (Eq.~\ref{CT}) to the tof measurements of Canali~\citep{canali1971drift} for <111> silicon.
 Selected results are shown in Fig.~\ref{fig:1/MU}, where the inverse mobility, $1/\mu (E) $, is plotted versus electric field.
 The constant low field mobility is clearly visible, with a sudden transition to a linear increase of $1/\mu $ with electric field, which occurs around $\unit[1.8]{kV/cm}$ for electrons and $\unit[2.5]{kV/cm}$ for holes at $\unit[300]{K}$.
 The transition point decreases with decreasing temperature.
 From the lower part of Fig.~\ref{fig:1/MU}, which shows the ratio of the fit results to the data, we observe systematic deviations between the data and both fits.
 However, the new KS parametrization shows somewhat smaller deviations, in particular at the transition between low- and high-field mobilities.

 To quantify the difference between the fits and the data, we use the $rdev$ (Eq.~\ref{eq:rdev}).
 The values of $rdev$ are $\unit[0.22\,(0.17)]{\%}$ for electrons, and $\unit[0.38\,(0.28)]{\%}$ for holes at $\unit[T=245\,(300)]{K}$ for the KS~fit.
 For the CT~fit they are $\unit[0.40\,(0.41)]{\%}$ for electrons and $\unit[0.47\,(0.42)]{\%}$ for holes.
 We conclude that the KS~parametrization provides a somewhat better description of Canali data for <111> silicon than the standard CT~parametrization.

 \subsection{Appendix 2: Current transients for the p-type sensor}
  \label{subsect:Appendix_p}

 \begin{figure}[!ht]
  \centering
   \begin{subfigure}[a]{7.5cm}
    \includegraphics[width=7.5cm]{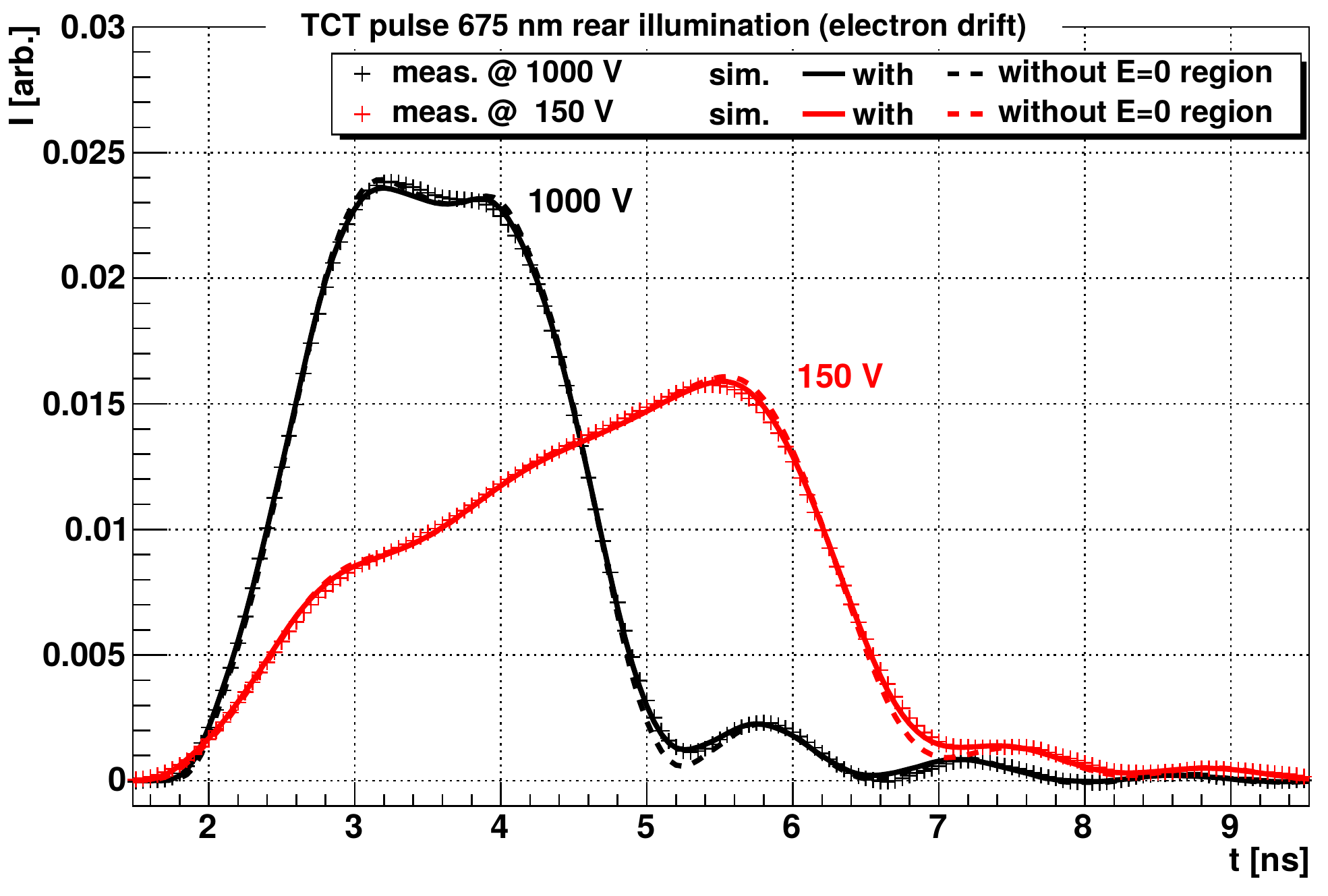}
    \caption{ }
     \label{fig:P_ElectronDrift}
   \end{subfigure}%
    ~
   \begin{subfigure}[a]{7.5cm}
    \includegraphics[width=7.5cm]{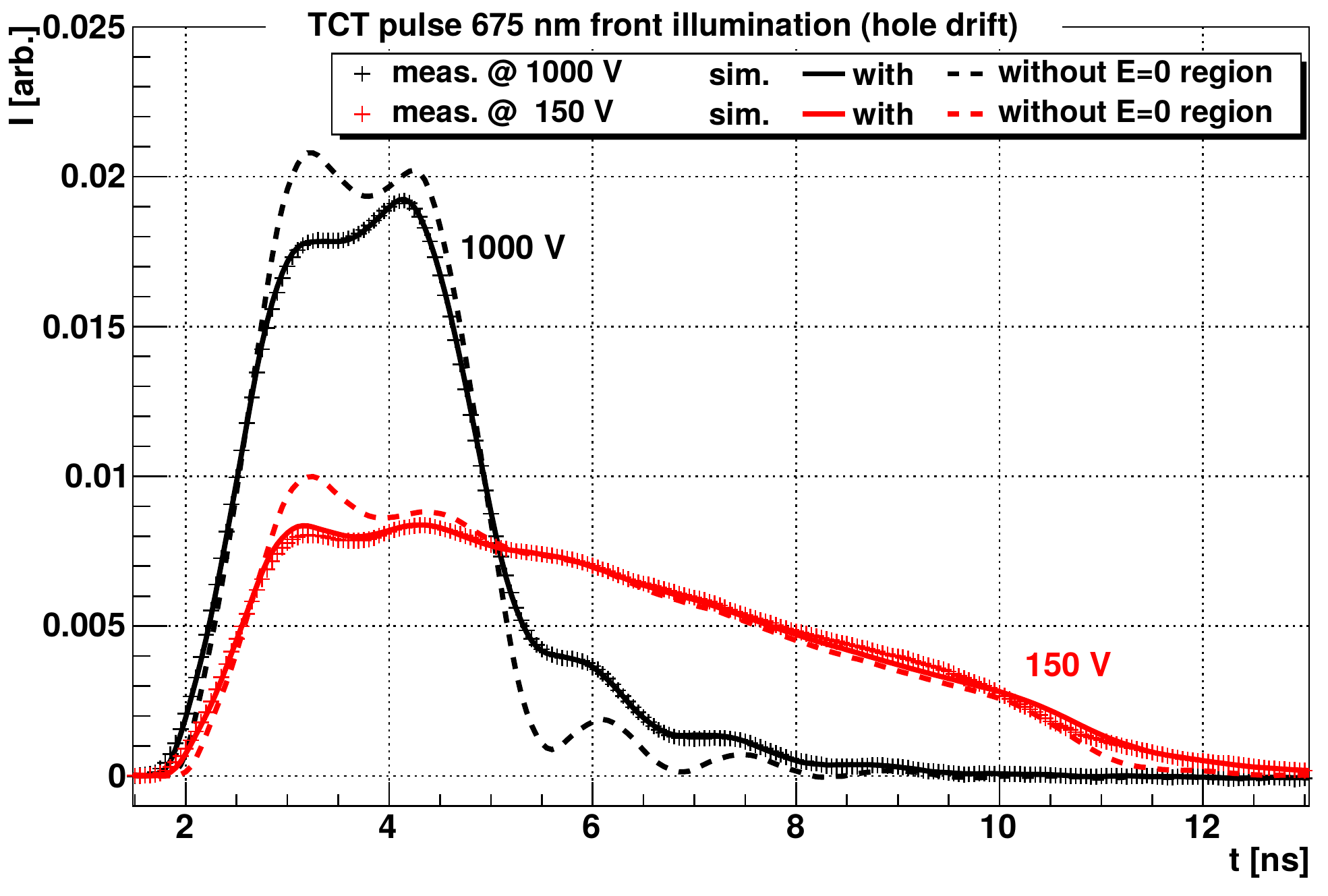}
    \caption{ }
     \label{fig:P_HoleDrift}
    \end{subfigure}%
   \caption{
  Comparison of the current transients of the p-type HPK diode measured with the CERN-SSD TCT setup using $\unit[675]{nm}$ laser light at $\unit[293]{K}$ (crosses) with the two simulations (lines).
  Solid lines: field-free regions of $\unit[1.6]{\upmu m}$ at the n$^+$p side and of $\unit[1]{\upmu m}$ at the p$^+$p side.
  Dashed lines: without field-free regions.
  The results at $\unit[150]{V}$ are shown in red, and the ones at $\unit[1000]{V}$ in black.
  (a) Rear-side illumination with the laser light, and
  (b) front-side illumination.
   }
  \label{fig:P_Drift}
 \end{figure}

 When performing the fits we were unable to achieve a good description of the current transients of the HPK p-type diode when we assumed an electric field, corresponding to a uniform p doping in the entire volume of the diode.
 However, adding a field-free region of $\unit[1.6]{\upmu m}$ at the n$^+$p side and one of $\unit[1]{\upmu m}$ at the p$^+$p side, resulted in a good description.
 To check the results we performed also measurements with the TCT setup at CERN-SSD Lab.
 It uses a wide-band amplifier and an Agilent scope with $\unit[2.5]{GHz}$ bandwidth and a sampling rate of $\unit[20]{GS/s}$.
 Using the Fast-Fourier-Transform method we successfully determined the transfer function, and performed fits to the measured current transients.
 Fig.\,\ref{fig:P_Drift} shows the results for the HPK diode at 150 and $\unit[1000]{V}$ measured at $\unit[293]{K}$.
 The data confirm the conclusions from the Hamburg measurements that field-free regions close to the diode surfaces have to be introduced in the simulation of the n$^+$p diode to obtain a satisfactory description.
 This is particularly the case for the front-side illumination shown in Fig.\,\ref{fig:P_HoleDrift}, where the signal is dominated by the drift of holes.
 Values in the range of 1 to  $\unit[2]{\upmu m}$, which result in a good description by the simulation, appear reasonable.
 However, we do not understand why a field-free region is required for the n$^+$p and not for the p$^+$n diodes.

\section*{Acknowledgment}
 \label{sect:Acknowledgement}

 The authors would like to thank Julian Becker for making available his current transient simulation program, Joern Schwandt for performing the TCAD simulations, and Eckhart Fretwurst, Erika Garutti and Joern Schwandt for stimulating discussions.
 We are also thankful to Michael Moll who gave us access to the CERN-SSD TCT setup, and to Hannes Neugebauer and Christian Gallrapp who helped in preparing and performing the measurements at CERN.
 The authors also thank the HGF Alliance \textit{Physics at the Terascale} for funding the Hamburg TCT setup.

\section*{References}
 \label{sect:References}

\bibliographystyle{unsrtnat}

\end{document}